\documentclass{egpubl}

\usepackage{eg2021}
 
 \ConferencePaper        
\usepackage[T1]{fontenc}
\usepackage{dfadobe}  

\usepackage{cite}  
\BibtexOrBiblatex
\electronicVersion
\PrintedOrElectronic
\ifpdf \usepackage[pdftex]{graphicx} \pdfcompresslevel=9
\else \usepackage[dvips]{graphicx} \fi

\usepackage{egweblnk} 

\usepackage{booktabs} 
\usepackage{xcolor}
\usepackage{multirow}

\usepackage{amsmath}
\usepackage{amssymb}
\usepackage{amsthm}
\DeclareMathOperator*{\argmax}{arg\,max}
\newtheorem{definition}{Definition}[section]

\usepackage{listings}
\title{Learning and Exploring Motor Skills with Spacetime Bounds}

 \author[L. Ma]{
   \parbox{\textwidth}{\centering 
   Li-Ke Ma\thanks{\texttt{milkpku@gmail.com}}$^{1,2}$  ~
   Zeshi Yang\thanks{\texttt{\{zeshiy,kkyin\}@sfu.ca}}$^1$ ~
   Xin Tong\thanks{\texttt{\{xtong,bainguo\}@microsoft.com}}$^3$~
   Baining Guo$^3$ ~
   KangKang Yin$^1$ 
   }
   \\
   \parbox{\textwidth}{\centering
   $^1$Simon Fraser University ~
   $^2$Tsinghua University ~ 
   $^3$Microsoft Research Asia}
 }

\begin{document}

\definecolor{myOrange}{rgb}{1,0.3,0}
\definecolor{myGreen}{rgb}{0,0.8,0.5}
\definecolor{kkBlue}{rgb}{0,0.0,0.9}

\newcommand{\like}[1]{\textcolor{myOrange}{#1}}
\newcommand{\zeshi}[1]{\textcolor{myGreen}{#1}}
\newcommand{\kk}[1]{\textcolor{kkBlue}{#1}}

\newcommand{\figref}[1]{Fig.~\ref{#1}}

\teaser{
 \includegraphics[width=0.75\linewidth]{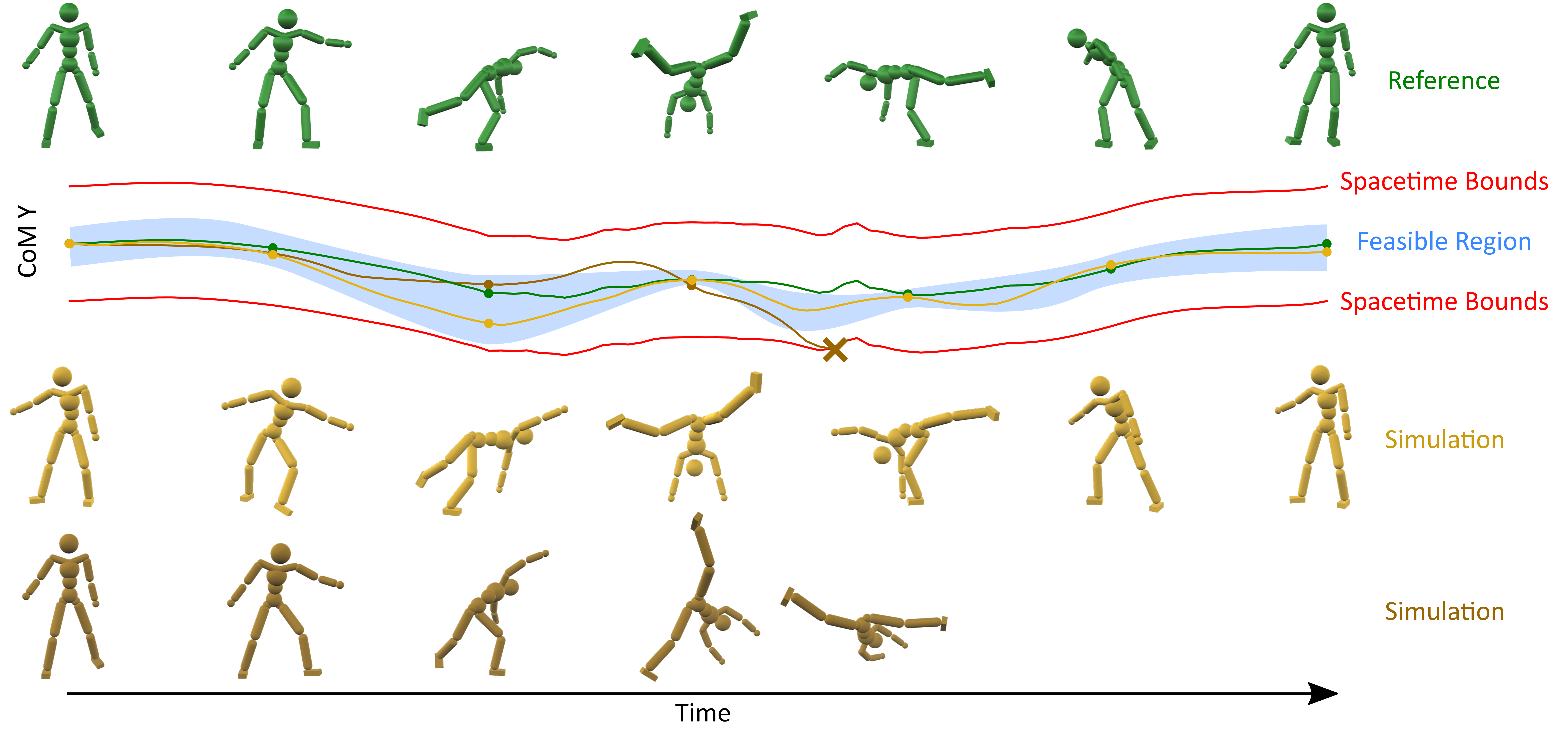}
 \centering
 \caption{Learning cartwheels with spacetime bounds. The top green motion shows the reference, and the bottom yellow motions are simulations. The curves represent the $Y$ position of the character's center of mass, and are colored to represent the reference (green), the simulations (yellow), and the spacetime bounds (red). The blue region illustrates the nonuniform feasible region under the given spacetime bounds. During training, episodes are terminated immediately once any spacetime bounds are violated, as shown in the bottom simulation.}
\label{fig:teaser}
}

\maketitle

\begin{abstract}
{Equipping characters with diverse motor skills is the current bottleneck of physics-based character animation. We propose a Deep Reinforcement Learning (DRL) framework that enables physics-based characters to learn and explore motor skills from reference motions. The key insight is to use loose space-time constraints, termed spacetime bounds, to limit the search space in an early termination fashion. As we only rely on the reference to specify loose spacetime bounds, our learning is more robust with respect to low quality references. Moreover, spacetime bounds are hard constraints that improve learning of challenging motion segments, which can be ignored by imitation-only learning. We compare our method with state-of-the-art tracking-based DRL methods. We also show how to guide style exploration within the proposed framework.}


\end{abstract}  

\section{Introduction}

{Recent years has seen many advances in physics-based character animation, especially since the application of Deep Reinforcement Learning (DRL) algorithms~\cite{Liu16,Peng:2018:DeepMimic,Yu:2018:LSLL,Park:2019:LPS}. These modern methods produce physically plausible motor skills either by tracking high quality reference motions~\cite{Peng:2018:DeepMimic,Park:2019:LPS}, or via smartly designed rewards~\cite{Yu:2018:LSLL}. However, tracking reference motions requires the existence of high quality example motions, and inherently prohibits any exploration of the potentially large feasible region of some motor skills for style variations. Designing good reward signals for DRL systems requires nontrivial domain knowledge and human insights, and does not directly support style exploration either.}

{We propose a simple DRL framework that can be used either standalone, or combined with imitation or hand-designed rewards. The proposed framework imposes spacetime constraints, hereafter referred to as \textit{spacetime bounds}, as they mainly bound the character states in space and time, during the reinforcement training process. That is, the DRL system only samples and accepts states within the spacetime bounds specified.}

{The advantages of the proposed bounding-constraint-based framework over a tracking-based system include:
\begin{itemize}
    \item Reward Simplification: Our framework can learn various motor skills with just a binary survival reward correlated to violations of spacetime bounds. An imitation or a hand-designed reward are not necessary anymore to reproduce motor skills. Simplified reward design and parameter tuning can potentially enhance a wider adoption of DRL methods in physics-based character animation.
    \item Increased Robustness: Tracking-based DRL methods stay close to a reference motion as much as possible. So when the reference motions are not in high quality, such as interpolated sparse keyframes, tracking methods may fail due to the physical implausibility of the reference motion. Spacetime bounds, however, are looser constraints. They allow freer exploration of the state space and thus may still succeed in finding physically plausible motions that resemble the low quality references. Meanwhile, spacetime bounds are hard constraints on the states. Challenging parts of reference motions, such as a quick 360$^{\circ}$  turn in a dynamic dance, can be ignored by an imitation reward to favor task success, but have to be respected when spacetime bounds are specified. Therefore our framework can reproduce skills more robustly and more faithfully.  
    \item Style Exploration: As spacetime bounds only loosely constrain the DRL exploration, multiple styles of a motor skill may be discovered. We show that using simple heuristic terms to reward metrics such as energy levels, different locomotion styles can be easily discovered. Such style exploration is quite challenging if possible at all using only imitation-based methods, as the style-encouraging terms conflict with reference tracking rewards.  
\end{itemize}
}

{We review the most relevant prior works in Section~\ref{sec:related_works}. Then we detail the concept of spacetime bounds and its interaction with the feasible region of the dynamic skill in Section~\ref{sec:motion_bound}. Our DRL training framework and important parameter setups are described in Section~\ref{sec:DRLTraining}. Various results are showcased in Section~\ref{sec:results}. Finally we conclude with discussions on limitations and future work in Section~\ref{sec:discussion}.}

\section{Related Works}
\label{sec:related_works}
{Synthesizing natural human motions in interesting styles has been a long-term central topic in character animation. Recently the machine learning community has also started to investigate generation of human-like motions using deep learning tools. Here we only review the work most relevant to ours.}

{\textbf{Kinematic models} can synthesize natural human movements with a pure data-driven approach \cite{Kovar:2002:MotionGraphs,Safonova:2007:OptimalSearch,Agrawal:2016:TaskBasedLocomotion,Clavet:2016:MotionMatching}. Latent representations can also be learned offline from data for synthesis at runtime \cite{Safonova:2004:OptPCA, Levine:2012:Embedding, Yongjoon:2010:MotionFields}. Recently, deep neural network are quite successful at achieving fast compact kinematic models that generalize better beyond the training data \cite{Holden16, Holden:2017:PFNN, Zhang:2018:MANN}. Such models can also serve as front-end motion generators for back-end physics-based models \cite{Bergamin:2019:DReCon, Park:2019:LPS, Won20}.} 

{Kinematic models have also been used extensively for motion style generation and transfer. The most direct approach is to extract style-related features from the style reference motion and then impose them on another content reference motion \cite{Amaya:1996:Emotion, Hsu:2005:ST, Shapiro:2006:SC, Ikemoto:2009:GME, Xia:2015:RST, Yumer:2016:spectral, Holden16}. 
In particular, \cite{Holden16} computes the Gram matrix of motion features extracted by an autoencoder to represent motion styles. Another approach is to parameterize motion variations into multiple factors. Statistic models are usually used to learn such factors from data \cite{Brand:2000:SM, Wang:2007:MGP, Min:2010:SEP}. In \cite{Aristidou:2017:ECU}, features related to emotions are mapped into 2D emotion coordinates to support motion editing of emotional styles. Our style exploration is inspired by these previous works, but we handle style exploration in a physics-based framework.}

{\textbf{Physics-based models} guarantee physical plausibility of synthesized motions, but are usually hard to design or learn. Robust locomotion controllers can be manually designed and tuned \cite{Yin03, Yin07, LeeYS10, Coros10, Coros:2011:LocomotionQuadrupeds}. When reference motions are available, trajectory optimization \cite{AlBorno13} or sampling-based controllers \cite{Liu:2010:Samcon, Liu:2015:Samcon2} can reconstruct open-loop controls that reproduce reference skills with high fidelity. Closed-loop controls that respond to perturbations can be constructed through model-predictive control methods \cite{Silva:2008:ModelPredictive, Hamalainen:2015:Online} or learned feedback mechanisms \cite{Ye10, Ding15, liu2012terrain, Liu16}.}  

{Tracking example motions, however, does not work on low quality references and prohibits exploration of new motion styles, as we will show in this paper. Our work is directly inspired by prior works that impose motion constraints rather than tracking references \cite{Agrawal:2013:Diverse, Liu:2002:SCDC}. \cite{Liu:2002:SCDC} transforms input sketches to physically plausible motions by detecting and imposing environmental constraints. \cite{Agrawal:2013:Diverse} encourages motion variations with respect to hand-crafted goal constraints. Our method imposes spacetime constraints during deep reinforcement learning, thus enabling robust learning and exploration of a diverse set of motor skills.}

{\textbf{Deep Reinforcement Learning} is a relatively new and effective approach to learning physics-based motor skills~\cite{lillicrap2015continuous, Heess:2017:Emergence}. The \textit{de novo} methods synthesize motor skills from scratch, and usually generate unnatural jerky motions. Special terms designed to lower energy and improve symmetry can be added to acquire more natural skills~\cite{Yu:2018:LSLL}. When kinematic reference motions are available, a more effective approach is to add an imitation reward term to encourage tracking of the reference \cite{2017-TOG-deepLoco,Peng:2018:DeepMimic,Peng:2018:SFV,Bergamin:2019:DReCon,Won20}. Such methods can usually synthesize high quality motor skills that are indistinguishable from the reference. The imitation reward, however, prohibits exploration of more stylized skills that are different from the reference. We replace reference tracking with spacetime bounds that are more supportive to style exploration. In addition, spacetime bounds also improve the learning robustness when the reference motion quality is low.}

{We show direct comparisons with DeepMimic \cite{Peng:2018:DeepMimic} in our results section. \cite{Won20} improves DeepMimic with multiplicative rewards and an early termination scheme based on reward values. Therefore, it can learn more diverse and robust skills than DeepMimic. Still, we show that our framework is more robust than \cite{Won20} for challenging skills, such as the ``mickey surprised'' motion as shown in our supplemental video. Furthermore, spacetime bounds are easier to specify than thresholds on rewards, and support style exploration better.}


\section{Spacetime Bounds}
\label{sec:motion_bound}

{Reference tracking-based DRL methods evaluate the quality of motor skills based on rewards that are real numbers. On one hand, it is hard to tell if a physics-based skill is successful or optimal from these numbers. {On the other hand, such rewards only encourage but do not guarantee  similarity between the learned skill and the reference motion.} For example, local joint angles or some portion of the skills may resemble the reference, but not the overall behavior or the full course of the motion.}

{We propose to constrain DRL learning with spacetime bounds instead. Spacetime bounds are constraints in space and time that correspond well with intuitive definitions of motor tasks. For example, a jump is a motor skill that for some duration of the motion both feet should leave the ground, and at no time of the motion should the character fall. \like{} In order to derive and analyze spacetime bounds, we start with the following definitions:
\begin{itemize}
    \item The \textit{state space}, denoted as $\mathcal{S}$, is the space of all possible states of a dynamical system.
    \item An \textit{event} is a state-time tuple $(s, t)$, and represents the system in state $s$ at time $t$. 
    \item A \textit{Spacetime}, denoted as  $\mathcal{M}$, is the space of all possible events.
    \item A \textit{trajectory} is a sequence of events in spacetime $\mathcal{M}$. {A trajectory is \textit{causal} if all points on the trajectory obey applicable physical laws and dynamic constraints.}
\end{itemize}
}


\vspace{\baselineskip}
\begin{definition}
A \textit{\textbf{spacetime bound}} $b$ is a subset of spacetime $\mathcal{M}$. We denote a set of spacetime bounds as $\mathcal{B}$.
\end{definition}

\subsection{Feasible Regions and Spacetime Bounds}
\label{ssec:feasible_region}

\begin{figure}
    \centering
     \includegraphics[width=\linewidth]{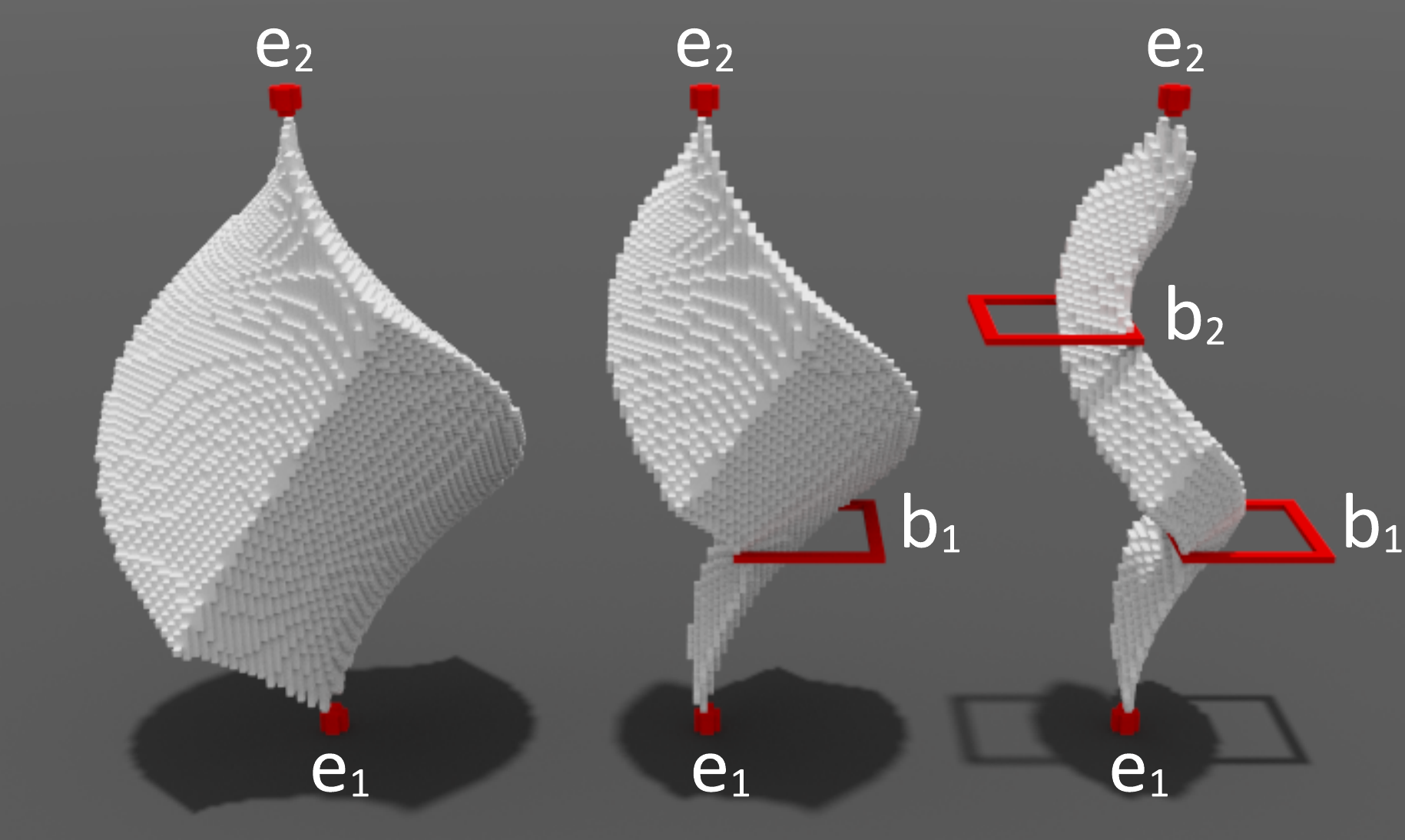}
    \caption{Feasible regions for a toy problem when more and more spacetime bounds are specified. Left: $\mathcal{B} = \{\{e_1\}, \{e_2\}\}$; middle: $\mathcal{B} = \{\{e_1\}, \{e_2\}, b_1\}$; right: $\mathcal{B} = \{\{e_1\}, \{e_2\}, b_1, b_2\}$. Note that in the right most case, the feasible region is much thinner than the specified red squarish spacetime bounds.}
    \label{fig:constraints_cases}
\end{figure}

{Spacetime bounds influence control learning via shaping the feasible region of a motor skill performed by a dynamical system.

\vspace{\baselineskip}

\begin{definition} 
The \textit{\textbf{feasible region}} associated with spacetime bounds $\mathcal{B}$, denoted as $Feasible(\mathcal{B})$, is the set of points on causal trajectories that $\mathcal{B}$ encloses. 
\end{definition}}


We first illustrate the influence of spacetime bounds on feasible regions using a toy problem. We restrict the motion of a mass point to the $X$ axis. So its state can be fully described by $(x, v)$, where $x$ is the position and $v$ is the velocity of the mass point. We then add forces to accelerate the mass point, but cap the acceleration at $a_{max}=2 m/s^2$. In Figure~\ref{fig:constraints_cases}, we visualize the feasible regions that correspond to more and more imposed spacetime bounds. Events $e_1=(x, v, t)=(0, 0, 0)$ and $e_2=(x, v, t)=(0, 0, 5)$ are initially specified as the start and end of a causal trajectory. $Feasible(\mathcal{B})$ where $\mathcal{B} = \{\{e_1\}, \{e_2\}\}$ directly reflects the amount of trajectories that connect $e_1$ and $e_2$. We then impose more and more spacetime bounds to shrink the feasible region. For example, two spacetime bounds 
\begin{equation}
  \begin{split}
    b_1 &= \{(x, v, T_1) | x \in [0.5, 2.5], v \in [-1, 1], T_1=5/3\} ~ and \\
    b_2 &= \{(x, v, T_2) | x \in [-2.5, -0.5], v \in [-1, 1], T_2=10/3\}
  \end{split}
\end{equation}
are added to generate the middle and right feasible regions in Figure~\ref{fig:constraints_cases}. We can see a significant shrinkage of the feasible region with each added spacetime bound.

\begin{figure}[t]
    \centering
    \includegraphics[width=\linewidth]{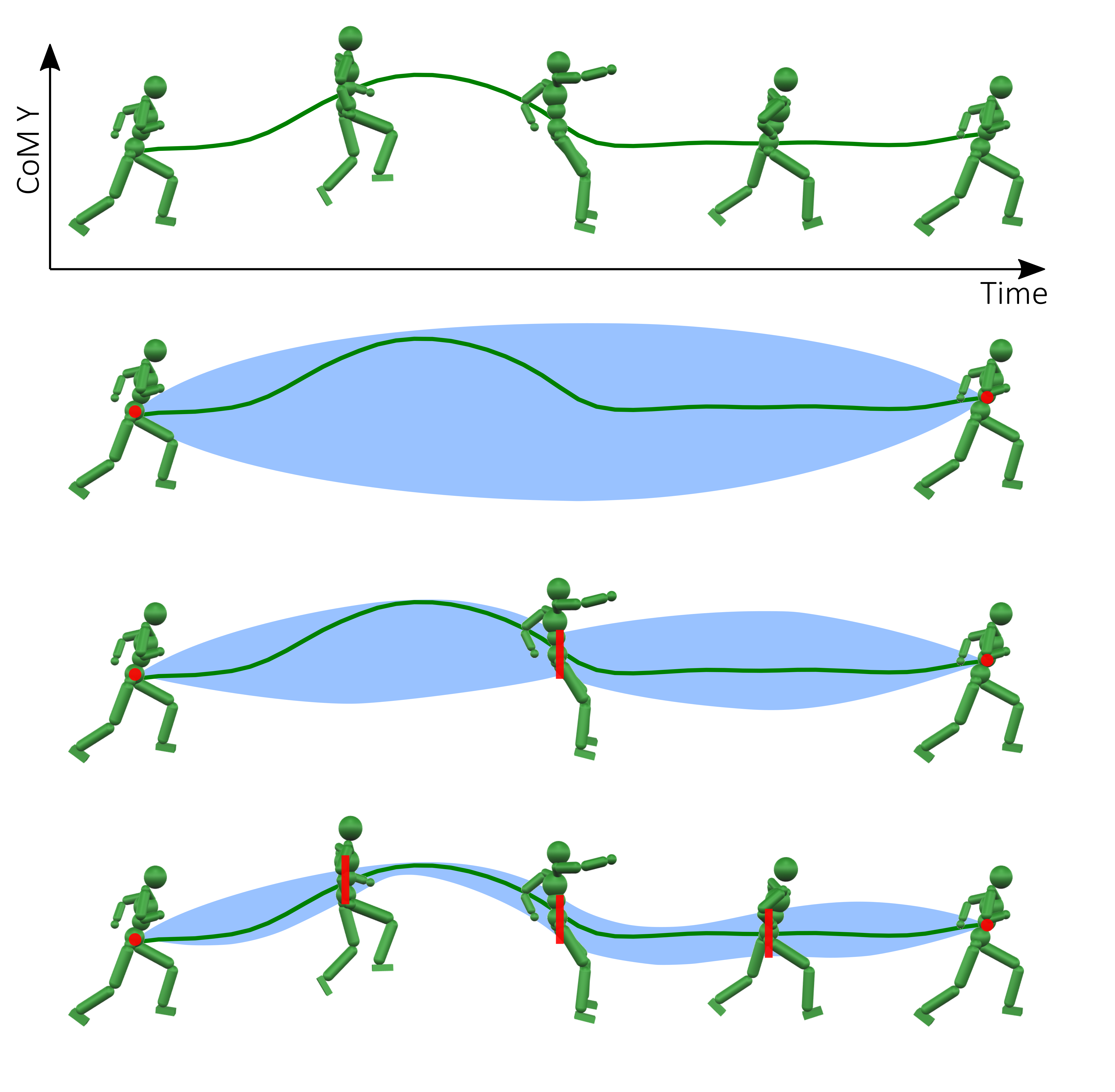}
    \caption{{Feasible regions of a run jump skill when more and more spacetime bounds are specified. Green curves are the character's CoM Y positions in the reference motion. Red dots and bars represent spacetime bounds. Blue regions indicate the feasible regions associated with the specified spacetime bounds. {Note that this is only a conceptual illustration. For all our results reported in Section~\ref{sec:results}, spacetime bounds are directly derived from the reference motion and applied to every frame of the simulation.}}}
    \label{fig:run_jump_feasible_region}
\end{figure}

{In more complicated systems such as a human-like character, the Degrees of Freedom (DoFs) are much larger and the system dynamics are much more complicated. We expect the feasible regions to shrink even faster. We illustrates such interaction between spacetime bounds and feasible regions in Figure \ref{fig:run_jump_feasible_region} for a run jump skill. Note that this figure is only conceptual and not mathematically accurate as in Figure~\ref{fig:constraints_cases}.}

{Generally speaking, different motor tasks performed by different dynamical systems have different intrinsic difficulties, which correlate to the volume of their feasible regions. For example, highly dynamic skills, such as a gymnastic backflip, are usually highly constrained. Their intrinsic feasible regions are usually quite small to start with. Consequently only professional athletes can perform such difficult motions in some optimal way. Low dynamic under-constrained motions, such as normal walking on flat ground, usually have larger initial feasible regions to start with. As a result, normal people can locomote and even in different styles. In either case, feasible regions shrink rapidly when more and more spacetime bounds are imposed. Therefore we do not need to specify spacetime bounds precisely or tightly for control policy learning.}

\subsection{Policy Learning with Spacetime Bounds}
\label{sec:motion_bound:method}

{Within a DRL framework, we sample an initial event $e$ in $\mathcal{M}$ and run the current policy $\pi_{\theta}$ to generate a trajectory as long as it stays inside $\mathcal{B}$. Once the trajectory violates $\mathcal{B}$, we terminate the current episode immediately. If the learning converges, then the final optimal controller $\pi_{\theta^*}$ can guarantee to generate trajectories within $Feasible(\mathcal{B})$.}


{We can construct spacetime bounds from the reference trajectory $m(t)$. More specifically, we define a spacetime bound at $t$ by restricting the state of the character to be within a region of size $\sigma$ centered at $m(t)$. For example, the root orientation should be within $50^\circ$ to the reference angle; or the end-effector positions should be within $0.5m$ to the reference positions. We set 
\begin{equation}
    \mathcal{B} = \{m(t, \sigma) | t \in [0, T]\},
\label{eq:training:restrictSpace}
\end{equation}
where $m(t, 0)$ is exactly the reference trajectory, and $m(t, \sigma)$ is the spacetime bound of size $\sigma$ at time $t$. $\sigma$ can be set uniformly for the whole duration of the motion, or as a function of time for finer control of the feasible region.} 

{At the beginning of learning, the policy is bad and therefore the trajectories violate the spacetime bounds very fast. This causes early termination of the training episodes. As the policy improves, training episodes will automatically become longer and longer. We thus do not need to employ a time-based curriculum strategy as in DeepMimic~\cite{Peng:2018:DeepMimic}, where episodes are heuristically scheduled to run longer and longer.}

{For motions that are highly constrained or in unstable equilibrium, the intrinsic corresponding feasible regions are narrow so loose spacetime bounds can already result in good controllers. For example, for a cartwheel we only need to bound its CoM positions (within $0.3m$) and orientations (within $40^\circ$). While for motions like locomotion, the initial feasible regions are relatively large so we need to tighten the spacetime bounds to learn skills that can reproduce the reference in a high fidelity. Alternatively, we can employ appropriate reward terms to guide the policy learning for more stylish skills.} 

{We note that special cases of the spacetime bounds have been used before. For example, DeepMimic~\cite{Peng:2018:DeepMimic} employs an early termination scheme that terminates an episode whenever certain links, such as the torso, make contact with the ground. This is equivalent to the following spacetime bound: 
\begin{equation}
    \mathcal{B} = \{ b_t | b_t = \{(s, t) | s\in S_{noUndesiredContacts}\}\}.
    \label{eq:training:deepmimic}
\end{equation}
Our spacetime bounds, however, are more general and customizable. These bounds impose stronger constraints than the DeepMimic early termination scheme alone, and thus greatly improve the sampling efficiency by not wasting time on large unrecoverable regions of the state space. Our spacetime bounds are also more flexible than imitation rewards, so that style exploration is possible during policy learning as we discuss next.}

\subsection{Style Exploration}
\label{sec:styleExploration}

{For motor skills with large default feasible regions, we can use style-related reward terms to explore different motion styles during DRL training. This is possible within our framework using spacetime bounds, as the bounds we specify are generally loose. In contrast, style exploration would be hard, if possible at all, using tracking-based methods, as the imitation reward and the style reward may conflict with each other or hard to tune.}

{\textbf{Heuristic Style Reward} -- We first achieve style exploration with two heuristic reward terms:
\begin{itemize}
    \item Kinematic Energy: We denote the kinematic energy calculated in the local frame defined at the CoM as $E$. Then the style reward for discovering motions at various energy levels is \begin{equation}
    r_s= 
    \begin{cases} 
        clamp(\frac{E_{max} - E}{E_{max} - E_{min}}, 0, 1), &\hspace{5pt}\text{to decrease energy}\\
        \\
        clamp(\frac{E - E_{min}}{E_{max} - E_{min}}, 0, 1), &\hspace{5pt}\text{to increase energy}
    \end{cases}
    \label{equ:kinematic_energy}
\end{equation}
where $[E_{min}, E_{max}]$ is the range where kinematic energy is linearly rewarded.  
    \item Volume: We denote the convex hull volume of selected points on the character as $V$~\cite{Aristidou:2017:ECU}. Then the style reward for discovering motions that span various volumes is
\begin{equation}
    r_s= 
    \begin{cases} 
        e^{-\frac{V}{\alpha}}, &\hspace{20pt}\text{to decrease volume}\\
        1 - e^{-\frac{V}{\alpha}}, &\hspace{20pt}\text{to increase volume}
    \end{cases}
    \label{equ:volume}
\end{equation}
where $\alpha$ is a scale parameter.
\end{itemize}}

{\textbf{Data-driven Style Reward} -- We also illustrate style exploration with the data-driven style term described in~\cite{Holden16}, where the style of a motion is encoded by the Gram matrix of its features extracted from a deep autoencoder $\Phi$. We directly use the autoencoder $\Phi$ trained from locomotion data in \cite{Holden16} for our experiments:
\begin{equation}
    r_s = e^{-\frac{\|\mathbf{G}_s - \mathbf{G} \|_2}{\alpha}},
    \label{equ:gram_matrix}
\end{equation}
where $\mathbf{G}_s$ is the Gram matrix of the motion in a desired style, and $\mathbf{G}$ is the Gram matrix of the simulated motion. We can then use stylized locomotion as our style descriptor to encourage the training to acquire policies that produce locomotion in similar styles. We also employ a regularization term to penalize large kinematic energy, large body linear accelerations and large joint angular accelerations, which may occur during style exploration with loose spacetime bounds. The regularization term is defined as:
\begin{equation}
    r_{reg} = e^{-\sum_{i=1}^{N} w_i a_i / \beta_i},
\end{equation}
where $a_i$ is the total kinematic energy, body linear accelerations, or joint angular accelerations. $\beta_i$ is a scale factor and $w_i$ is a weight that sums up to 1 for $i \in \{1,\cdots,N\} $. $N$ is $1$ plus the number of body parts and the number of joints. The final reward is thus 
\begin{equation}
    r = r_s \cdot r_{reg}.
\end{equation}
}

{The heuristic and data-driven style rewards as defined above are simple to implement and effective in discovering interesting motion styles as will be shown in Section~\ref{ssec:style_discovery}.} 
\section{DRL System}
\label{sec:DRLTraining}




{Reinforcement learning of motor skills is formulated as a Markov Decision Process (MDP). The goal is to learn a policy $\pi^*$ that maximizes the expected long-term reward:
\begin{equation}
      \pi^*= \argmax_{\pi} \mathbb{E}_{s \sim \rho_0} \big[V_{\pi}(s)\big],
\end{equation}
where the policy $\pi$ outputs a distribution of actions $a_t\sim\pi(a|s_t)$ when given a state $s_t$, $\rho_0$ is the distribution of initial states. 
\begin{equation}
    V_{\pi}(s_t) = \mathbb{E}_{s_t, a_t \sim \pi} \big[\sum_{t=0}^{t=T} \gamma^t r_t\big]
\end{equation}
is the value function, which is the expected discounted cumulative reward of $\pi$ starting from state $s_t$. $T$ can be either finite or infinite, $r_t$ is the reward at time $t$, and $\gamma \in (0, 1)$ is the discount factor. We refer interested readers to \cite{Sutton:2018:reinforcement} for more theoretic derivations. We use an actor-critic DRL architecture and parameterize both the policy and the value function using deep neural networks. Similar to DeepMimic, we train the networks with a collection of algorithms such as PPO  ~\cite{Schulman:2017:PPO}, $TD(\lambda)$ and $GAE(\lambda)$~\cite{Schulman:2015:GAE}. We refer interested readers to \cite{Peng:2018:DeepMimic} for more detailed explanations. We terminate training episodes when the specified spacetime bounds are violated as shown in Figure~\ref{fig:teaser}, or when the end state is reached, or when the time limit is exceeded. }

\subsection{States and Actions}
\label{ssec:state_and_action}
{The system state constitutes a phase index $\phi$, and the position $p$, orientation $q$, linear velocity $v$, and angular velocity $\omega$ of each link, and the position of chosen end-effectors $p_e$. All kinematic quantities, except the root orientation, are calculated in a local frame attached to the root and aligned with the motion direction as described in~\cite{Ma:2019:TRD}. Therefore the state features are invariant to the motion direction, which is the $X$ axis by default. For motions that contain $360^\circ$ rotations in the sagittal plane, such as backflips and rolling, the $Z$ axis is selected as the motion direction.}


{Each internal joint is activated by a PD (Proportional Derivative) servo. The action vector therefore consists of target orientations for these PD controllers. Each rotational joint is either a 1-DoF revolute joint or a 3-DoF spherical joint. We choose to parameterize input orientations in quaternions, and target and output orientations in exponential maps \cite{Grassia:1998:ExpMap}, among the multiple choices for parameterizing 3D rotations \cite{Liu16, Holden:2017:PFNN,Peng:2018:DeepMimic,Yu:2018:LSLL, Bergamin:2019:DReCon,Park:2019:LPS}.} 


\subsection{Network Structure}
\label{ssec:network_structure}
{Figure~\ref{fig:network:fdm_actor} shows our policy network, which consists of an open-loop feedforward controller (FFC) and a feedback controller (FBC). The FFC looks up the kinematic reference motion and outputs the default target joint angles. The FBC is a trained neural network that outputs corrections to the FFC. This structure is inspired by previous works where controls are decomposed into a feedforward component and a feedback component. Such controllers are more robust and compliant in general, and faster to learn in DRL settings \cite{Yin03,Liu:2010:Samcon,Ding15,Bergamin:2019:DReCon}. More specifically, the FFC stores the joint angles from the reference and linearly interpolates them at run time according to the current phase index $\phi$ to generate $\hat{\mathbf{q}}$. The FBC consists of two fully-connected layers with 1024 and 512 hidden units respectively, and outputs correction angles $\Delta \mathbf{q}$. We use ReLU activation for each layer. The final output target angle for the PD servo is then $\mathbf{q} = \hat{\mathbf{q}} + \Delta \mathbf{q}$. All angles $\hat{\mathbf{q}}$, $\Delta \mathbf{q}$, and $\mathbf{q}$ are parameterized in exponential maps. {Alternatively quaternions could be used for the parameterization and quaternion multiplication could be used for the angle correction. Our learning framework can be applied just the same.}}

\like{
\begin{figure}[]
    \centering
    \includegraphics[width=\linewidth]{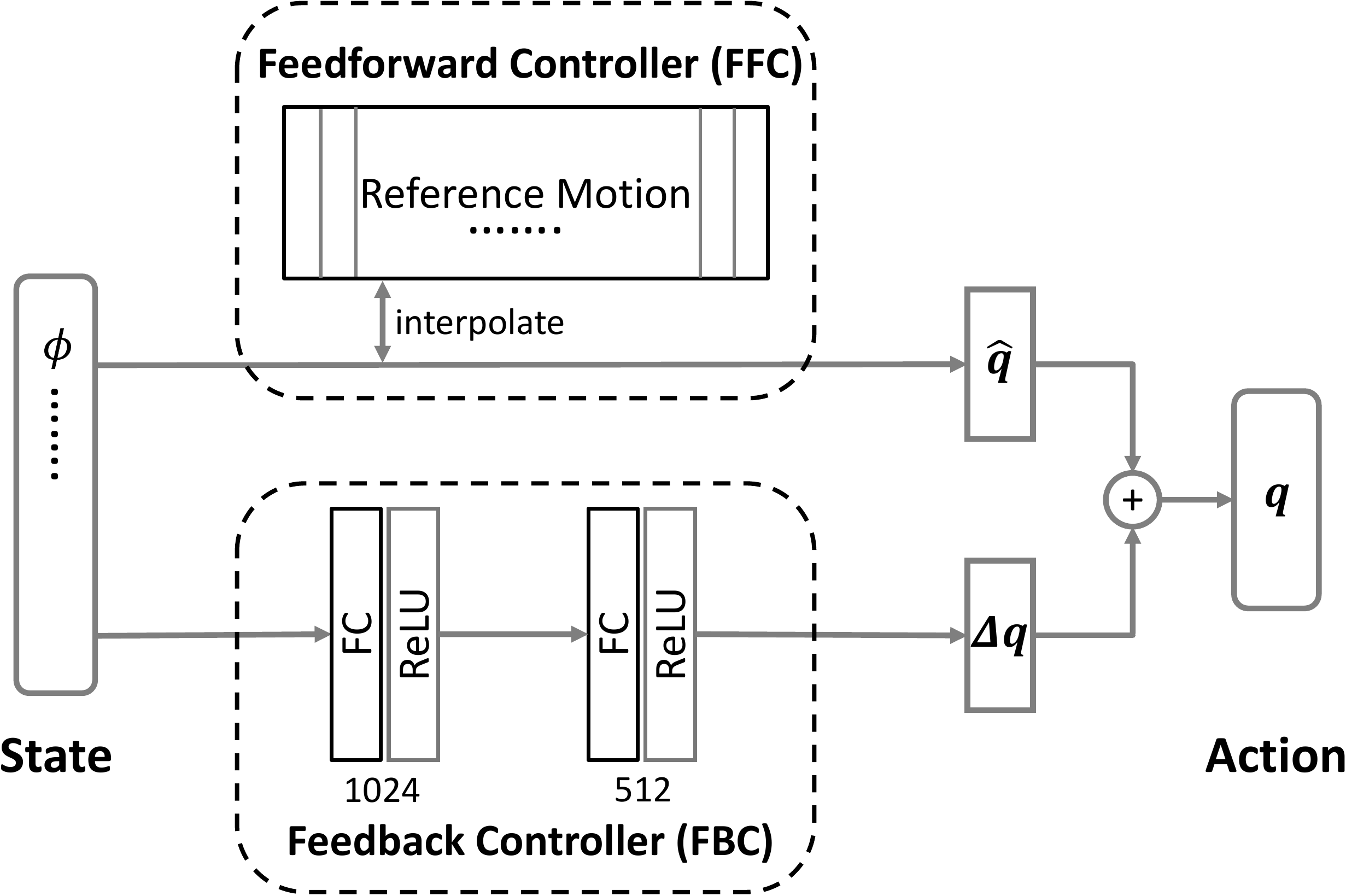}
    \caption{Our policy network consists of a feedforward controller (FFC) and a feedback controller (FBC). The FFC outputs default joint angles $\hat{\mathbf{q}}$ according to the phase index $\phi$ and the reference motion. The FBC is a two-layer fully-connected neural network with 1024 and 512 hidden units respectively. It takes in the full state vector and outputs offset joint angles $\Delta \mathbf{q}$. The final control signal $\mathbf{q} = \hat{\mathbf{q}} + \Delta \mathbf{q}$.}
    \label{fig:network:fdm_actor}
\end{figure}
}


{Our value network is similar to the feedback control branch of the policy network, except that the final output is a scalar that estimates the value function $V_{\pi}(s)$}.
  

\subsection{Initial States Adaptation}
\label{sec:training:adaptive_sampling}
{The initial state distribution $\rho_0$ determines the states in which an
agent begins each episode. Reference State Initialization (RSI) proposed in~\cite{Peng:2018:DeepMimic} has been proven to help the agent to access desirable states early in the learning, and thus improves the efficiency and robustness of DRL algorithms. In our framework, the RSI strategy is equivalent to setting the initial events to $E=m(t, 0)$ where $t$ is uniformly sampled. The RSI strategy does not work well for challenging skills or low-quality references, however. First, the feasible regions are not uniform in size across time. At critical points where the motions are more likely to fail, drawing more samples will likely help. Second, the sampled initial states could be infeasible from low-quality references. A strategy to help evolve the set of initial states into the feasible regions will be beneficial. We thus develop the following two adaptation strategies to further improve the robustness of learning: importance sampling from $E$, and evolving $E$ by training experiences.}

\subsubsection{Importance sampling of reference motion}
\label{sec:importance}
{Generally speaking, learning is more likely to fail early from initial states sampled around critical points of a motor skill. We therefore sample more around critical points of the reference motion. More specifically, we first uniformly divide the reference motion into $n$ segments. Denote the average estimated return starting from states in segment $k$ as $w_k$. Then for each new episode, we sample each segment with probability 
\begin{equation}
    p(k) = (1-u)\frac{\exp{(-w_k/v)}}{\sum_{i=1}^{n} \exp{(-w_i/v)}} + \frac{u}{n}, 
\end{equation}
where $u=0.2$ is the probability to sample uniformly for all our experiments, and $v$ is a scale parameter adaptively set to $(\max{w_k} - \min{w_k})/3$. This is similar to the adaptive sampling scheme described in \cite{Park:2019:LPS}. {There are also other adaptive schemes to facilitate learning in the literature, such as utilizing value functions to guide the sampling of body shapes  \cite{won2019learning}, or learning progressively from easier tasks to harder tasks {\cite{xie2020allsteps}} which is equivalent to the adaptive sampling proposed in \cite{Park:2019:LPS,won2019learning}.}

\subsubsection{Initial States Evolution}
\label{sec:evolution}
When the reference motion is of low quality, such as hand animation from sparse keyframes, sampled initial states may be outside of the feasible region. We develop a scheme to select elite states from experiences to gradually guide them into the feasible region. More specifically, we assign a buffer for each motion segment to hold the current set of elite initial states. These buffers are initialized with $m$ events sampled from the original reference motion segments. We then sample initial states from these buffers for training. After each epoch, we use the Boltzmann distribution to draw $m$ elite samples from all collected samples to overwrite the buffer:
\begin{equation}
  p(l) = \frac{\exp{(-w_l/v)}}{\sum_{i}\exp{(-w_i/v)}},
\end{equation}
where $w_l$ is the estimated return of the $l$th state in the buffer, and $v$ is a scale parameter that we set to $(\max{w_l} - \min{w_l})$. We note that the ASI (Adaptive State Initialization) scheme described in \cite{Peng:2018:SFV} and the CMA (Covariance Matrix Adaptation) scheme described in \cite{Liu:2015:Samcon2} share a similar motivation, but our scheme is much simpler to implement and works well for all the results shown in the paper.}


\section{Results}
\label{sec:results}

{We implemented our framework in PyTorch~\cite{pytorch} and Bullet \cite{Bullet}. Our character model weighs 45 $kg$, and has 15 internal joints and 34 DoFs in total. Each joint except for the root is actuated by a stable PD controller~\cite{Tan:2011:SPD}. We run the simulation at 600 $Hz$, and the control at 30 $Hz$.}

{For DRL training, we use a binary survival reward at each control step. If the state is within the spacetime bounds, the character earns a reward $1$, otherwise $0$ and the training episode is terminated immediately. When there are other rewards, such as style encouraging rewards, we simply multiply the binary survival reward with the other rewards. Since the survival reward is $1$, the reward value is simply the value of the other rewards. We set the reward discount factor $\gamma=0.95$, and $\lambda=0.95$ for both $TD(\lambda)$ and $GAE(\lambda)$. The learning rate is $2.5\times10^{-6}$ for the actor network and $1.0\times10^{-2}$ for the critic network. In each training epoch, we sample 4096 state-action tuples in parallel on multi-core processors. The training batch size is 256. We report the performance statistics on a desktop with an 18-core Intel i9-7980XE CPU, where training takes about 30 minutes to 24 hours, depending on the length and difficulty of the motor skills.}


{We first show that our method can train controllers performing basic tasks with only spacetime bounds in Section~\ref{ssec:reward_elimination}. Then in Section~\ref{ssec:robustness}, we present more challenging cases where reward-based methods fail but our method can still succeed. Next in Section~\ref{ssec:style_discovery}, we demonstrate a variety of motion styles synthesized by our method, either using heuristic or data-driven style rewards. Lastly in Section~\ref{ssec:ablation} we conduct ablation studies on the sensitivity of spacetime bounds, and the effect of FFC and the initial states adaptation. We also refer our readers to the supplemental video for visual assessment of our results.}


\subsection{Learning without Tracking}
\label{ssec:reward_elimination}

\begin{table}[t]
    \centering
    \begin{tabular}{|c|c|c|c|}
        \hline
        task      & $T_{cycle}(s)$ & Ours $N_s(10^6)$ & DeepMimic $N_s(10^6)$  \\
        \hline
        walk      & 1.26           & 4.08             & 23.80                  \\
        run       & 0.80           & 4.11             & 19.31                  \\
        jump      & 1.77           & 41.63            & 25.65                  \\
        roll      & 2.02           & 12.31            & 23.00                  \\
        cartwheel & 2.72           & 17.35            & 30.45                  \\
        dance     & 1.62           & 10.00            & 24.59                  \\
        run jump  & 1.53           & 11.02            & 24.07                  \\
        backflip  & 1.75           & 41.20            & 31.18                  \\
        \hline
    \end{tabular}
    \caption{Number of training samples $N_s$ needed for character to be able to perform tasks for 20 seconds without falling. }
    \label{tab:rwd_elimination}
\end{table}

\begin{figure*}[t]
    \centering
    \includegraphics[width=\linewidth]{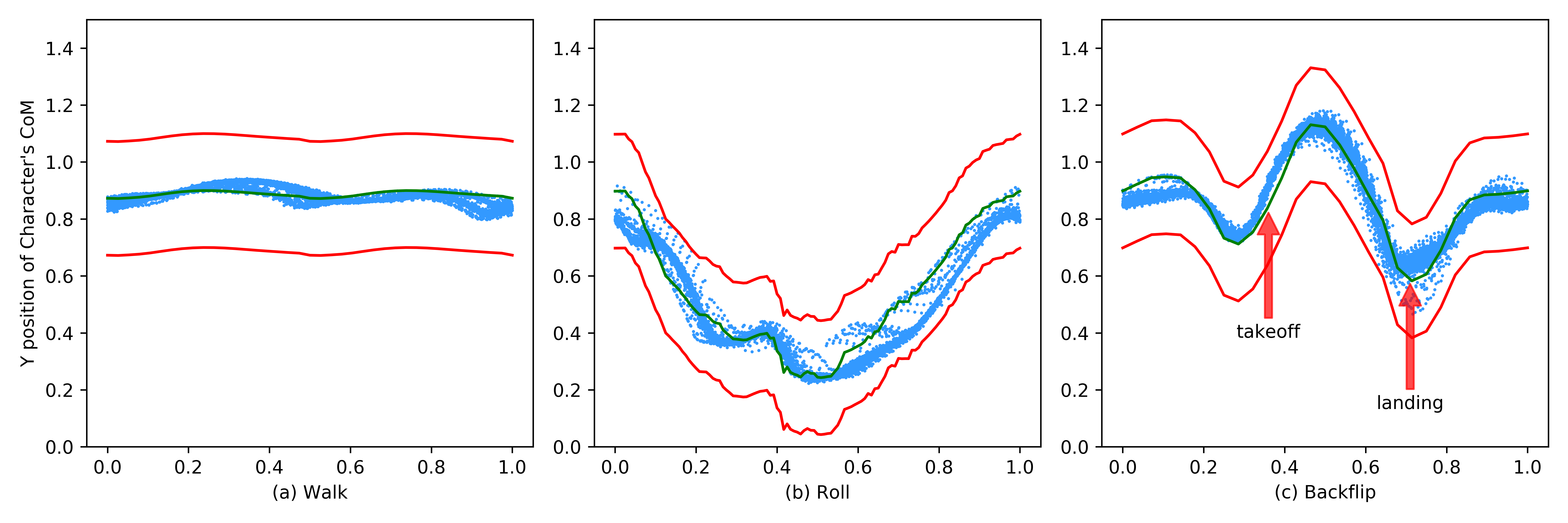}
    \caption{CoM $Y$ position with respect to the phase index. The green line is the reference and the red lines represent the spacetime bounds. Blue dots are samples from 100 episodes of the learned policy. We can see that the feasible regions of the learned skills lie inside the spacetime bounds; and their sizes are not uniform. For example in backflip (c), it is narrower around the takeoff and wider around the landing.}
    \label{fig:boa}
\end{figure*}

\begin{figure}[t]
    \centering
    \includegraphics[width=\linewidth]{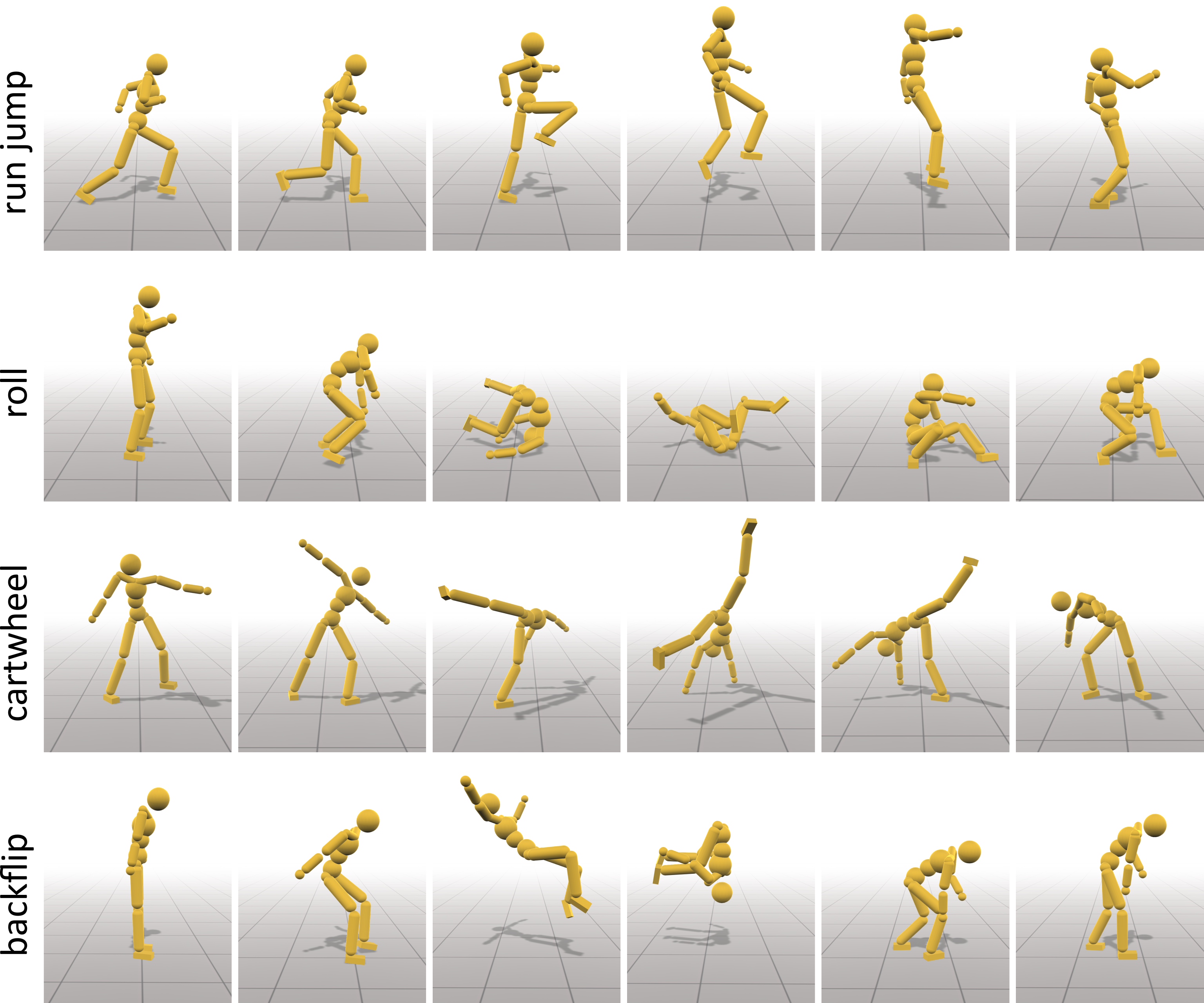}
    \caption{Cyclic skills trained with loose spacetime bounds and no imitation reward.}
    \label{fig:results:rwd_elimination}
\end{figure}

{We first train the character to follow reference motions without any imitation reward, but with loose spacetime bounds $m(t, \sigma)$ with $\sigma$ set as follows:
\begin{itemize}
    \item CoM $x,y,z$ positions: $0.2~m$
    \item Root and joint orientation: $0.7~rad \approx 40^\circ$
    \item Endeffector distance: $0.5~m$
\end{itemize}
CoM positions are compared in each dimension $x,y,z$ separately. Joint orientations are compared in their local frames. Endeffector distances are compared in a direction-invariant local frame, same as the one that we use to derive the state representations as described in Section~\ref{ssec:state_and_action}. Table~\ref{tab:rwd_elimination} lists the number of samples needed for learning each skill, and Figure~\ref{fig:results:rwd_elimination} shows the snapshots of learned skills. We note that not all spacetime bounds are needed for all skills. For highly dynamic motions such as the cartwheel, learning can be successful with just the bounds on CoM position and root orientation. {Generally speaking, spacetime bounds for COM position and root orientation bound the overall behavior of the character, such as moving forward or moving upward. Bounds on local joint orientations address the local pose similarity. Bounds on end-effectors prevent accumulated errors on a chain caused by individual joint angle deviations.} }

{Figure~\ref{fig:boa} illustrates the relationship between the reference motion and the spacetime bounds. It plots the CoM $Y$ position with respect to the phase index. The reference $m(t)$ is centered by the spacetime bounds $m(t, \sigma)$. As shown in Figure~\ref{fig:boa}(b), sampled CoM $Y$ positions from simulations controlled by the learned policy lie inside the spacetime bounds, and can notably deviate from the reference. Figure~\ref{fig:boa}(c) reveals that the feasible region is narrower around critical points such as the taking off phase of a backflip, and wider around stable regions such as the landing phase.}

\begin{figure*}[t]
    \centering
    \includegraphics[width=\linewidth]{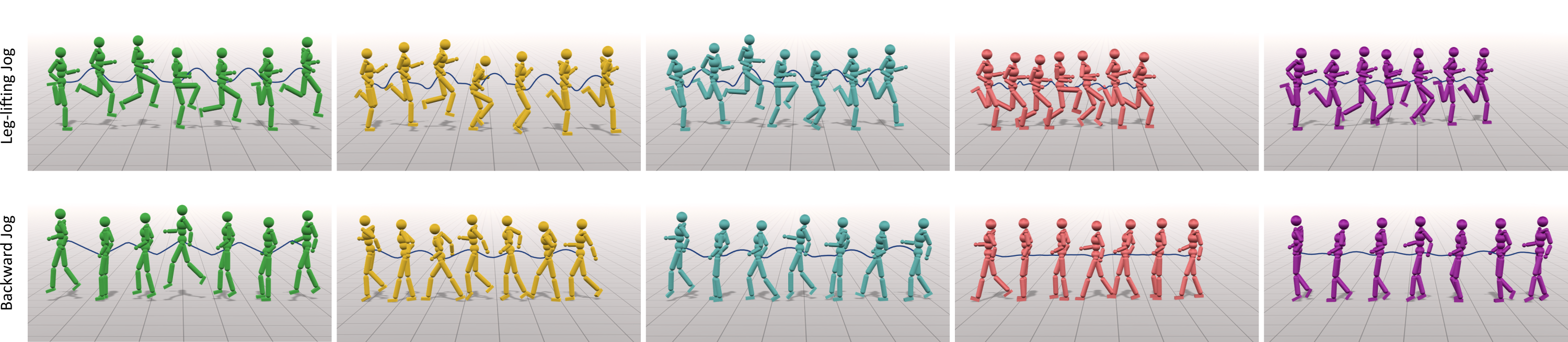}
    \caption{Robustness comparison between policies trained using our DRL framework with different options and DeepMimic, for low quality references from keyframed sparse poses. Green characters represent the reference motions. Yellow, blue, and red characters represent policies trained using our DRL framework with spacetime bounds only, with both spacetime bounds and imitation rewards, and with imitation rewards only. Purple characters represent policies trained using DeepMimic. For both skills, policies trained with spacetime bounds can better reproduce the intended skills.}
    \label{fig:robustness:low_quality}
\end{figure*}

\begin{figure}[t]
    \centering
    \includegraphics[width=\linewidth]{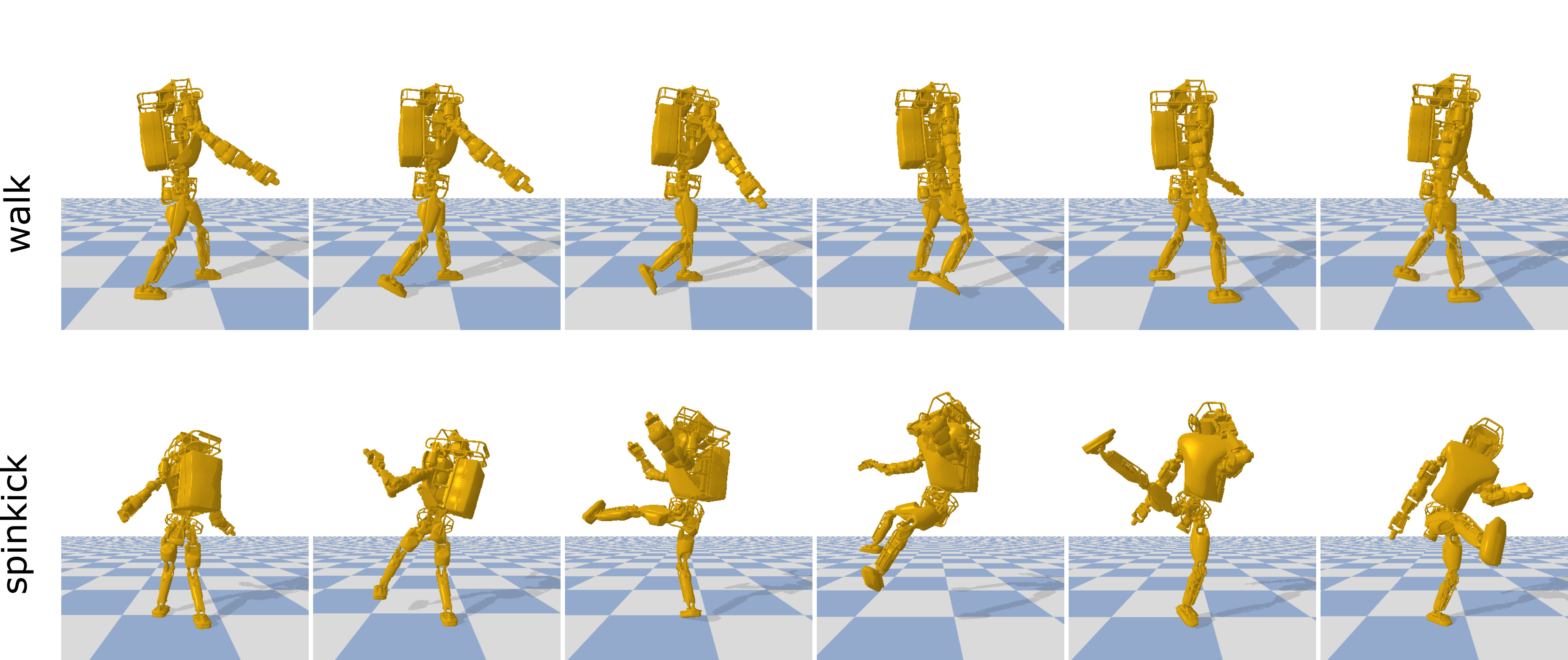}
    \caption{Retargeting human motions to an Atlas robot. Policies are trained using our DRL framework with spacetime bounds only.}
    \label{fig:robustness:retargeting}
\end{figure}

{Tuning spacetime bounds for internal joints is usually easy, as the reference can be tracked well through PD controllers, especially for those joints that do not support the body weight, such as upper body joints in locomotion skills. CoM positions and root orientations, however, are not directly actuated and controlled. Yet it is critical to follow them to achieve the desired motor skills. In addition, errors for the CoM horizontal position accumulate in time. Therefore, it takes some time to experiment proper spacetime bounds for the CoM and the root. Nevertheless, we are able to find one set of spacetime bounds for all the motions that we tested.}

\begin{figure}[t]
    \centering
    \includegraphics[width=\linewidth]{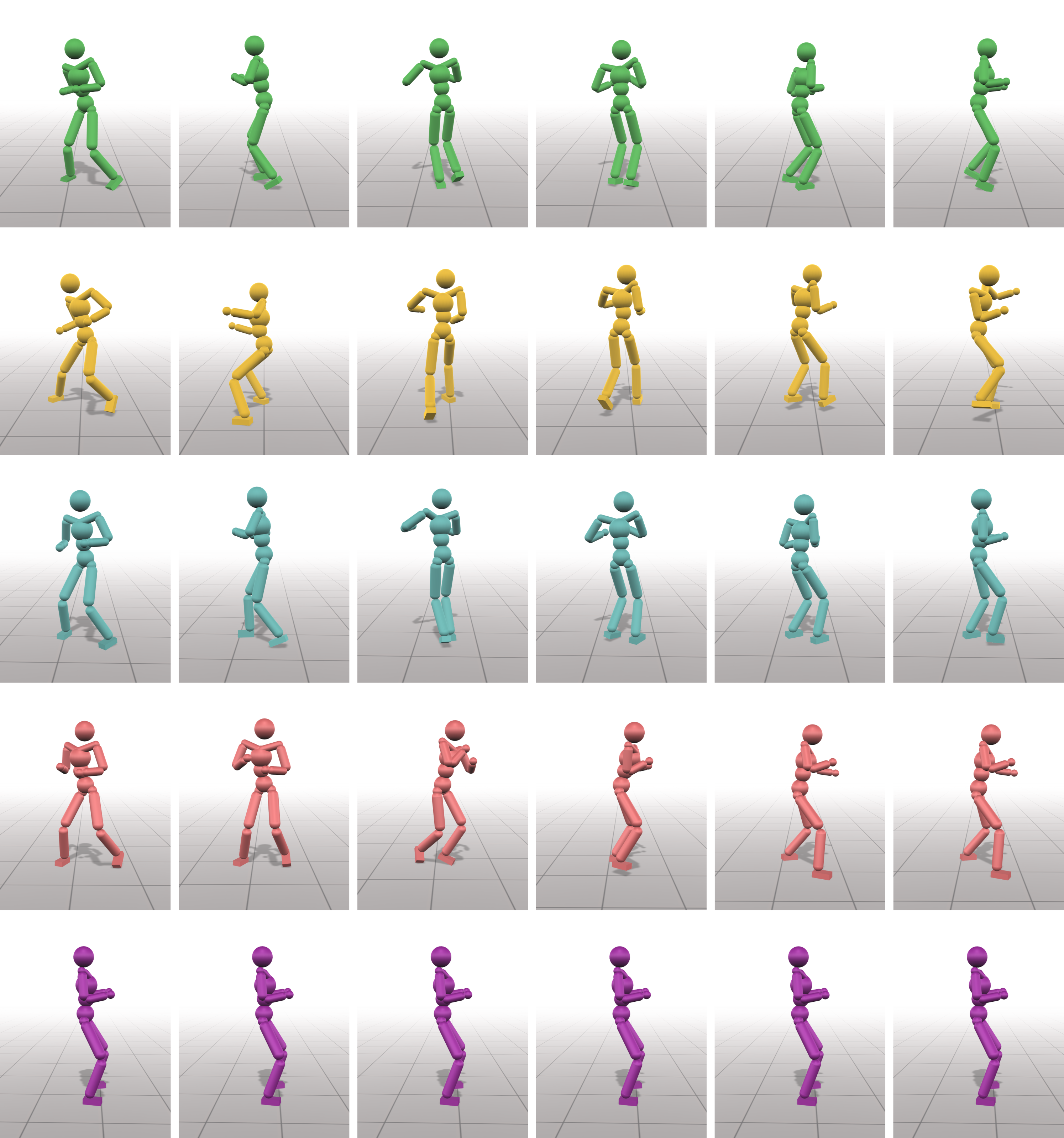}
    \caption{Robustness comparison between policies trained using our DRL framework with different options and DeepMimic, for a challenging skill break dance. Green characters represent the reference motions. Yellow, blue, and red characters represent policies trained using our DRL framework with spacetime bounds only, with both spacetime bounds and imitation rewards, and with imitation rewards only. Purple characters represent policies trained using DeepMimic. The reference character performs two 360$^\circ$ jump turns during the dance. The policies trained with spacetime bounds are able to reproduce the reference motion, but the one trained with imitation rewards only cannot reproduce the challenging parts. The policy trained using DeepMimic fails completely.}
    \label{fig:robustness:difficult}
\end{figure}

\subsection{Robustness to Challenging Cases}
\label{ssec:robustness}

{Tracking-based DRL systems may fail for difficult motor skills or due to poor reference quality. For example, when the reference motion contains fast body rotations or sophisticated foot work, the tracking-based DRL system tends to sacrifice tracking fidelity of these challenging motion segments in order to gain longer survival which leads to bigger total rewards. Our spacetime bounds, however, set hard boundaries for how much the learned skills can deviate from the reference, so the final motion will be guaranteed to be within a certain neighborhood of the original reference. On the other hand, our spacetime bounds treat all trajectories within the preset neighborhood equally, so that deviation from low quality reference is easier in order to achieve robust skills. In contrast, tracking-based methods have to compromise the quality of learned skills for more accurate tracking of the bad reference to gain more rewards.}

\begin{table}[t]
    \centering
    \begin{tabular}{|c|c|c|c|c|}
         \hline
         $w_s$                   &    0.5   &          0.6         &            0.7          & 0.8 \\
         \hline
         cartwheel $E\downarrow$ & no style &       unstable &  failed & failed \\
         cartwheel $V\downarrow$ & no style &       no style       &  unstable   & failed \\
         dance $E\downarrow$     & no style &   weird style   &  failed  & failed\\
         dance $E\uparrow$       & no style &     slight style     &  failed  & failed\\
         \hline
    \end{tabular}
    \caption{Style exploration using an imitation reward rather than spacetime bounds. $E\uparrow$ and $E\downarrow$ show motions with kinematic energy encouraged and discouraged respectively, and $V\uparrow$ and $V\downarrow$ show motions with full body volume encouraged and discouraged respectively. Most results are style-less motions, or unstable or failed skills.}
    \label{tab:sensitivity}
\end{table}

{We compare policies trained using our DRL framework with differnt options and the original DeepMimic, keeping same parameter settings wherever possible. Figure~\ref{fig:robustness:low_quality} shows two reference motions of rather low quality, keyframed from sparse jogging poses. For both tasks, policies trained with spacetime bounds can reproduce the reference tasks in similar styles, while policies trained without spacetime bounds or the original DeepMimic cannot. Figure~\ref{fig:robustness:difficult} shows that the policy trained with spacetime bounds can reproduce 360$^\circ$ jump turns in a break dance, but policies trained without spacetime bounds or DeepMimic cannot.} 

{We also perform retargeting experiments using our framework, as shown in Figure~\ref{fig:robustness:retargeting}. We use the built-in Atlas robot model from PyBullet~\cite{Bullet}. The morphology of this robot is significantly different from that of humans. The model also uses three revolute joints to model spherical joints, so we parameterize rotations of spherical joints, such as shoulders and hips, using Euler angles. We directly use the same motions captured from human performers as references without any kinematic retargeting. Our framework is able to physically retarget locomotion and gymnastics skills onto the robot model using spacetime bounds only.}



\begin{figure}[t]
    \centering
    \includegraphics[width=\linewidth]{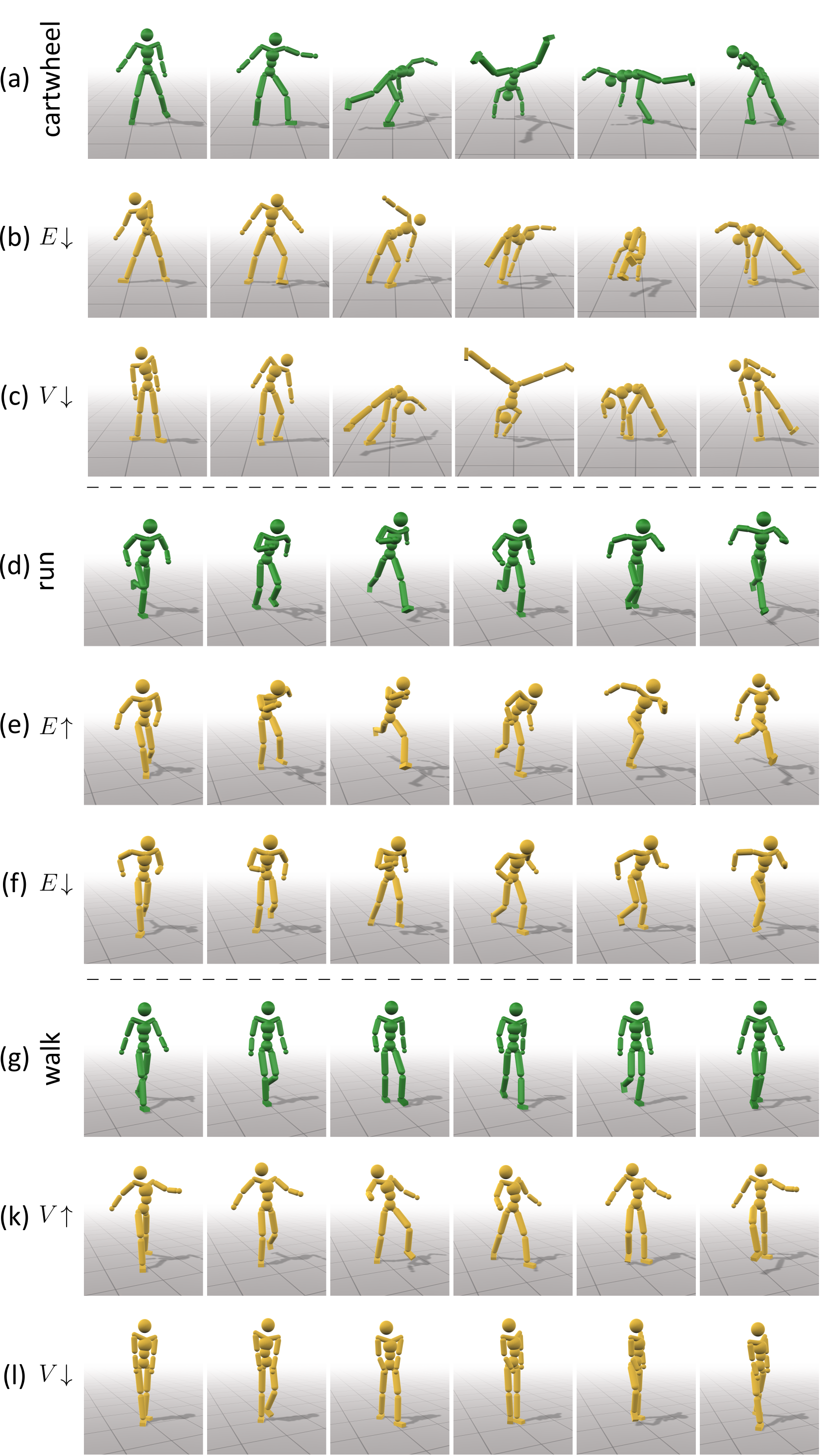}
    \caption{Styles explored by using heuristic style rewards with spacetime bounds. Reference motions are colored in green and stylized motions in yellow. $E\uparrow$ and $E\downarrow$ show motions with kinematic energy encouraged and discouraged respectively. $V\uparrow$ and $V\downarrow$ show motions with full body volume encouraged and discouraged.}
    \label{fig:new_style}
\end{figure}

\subsection{Style Exploration}
\label{ssec:style_discovery}

\subsubsection{Heuristic Styles}
\label{sssec:heuristic_styles}
{We combine spacetime bounds with the heuristic style rewards as described in Section~\ref{sec:styleExploration} to generate stylized motions, some of which are shown in Figure~\ref{fig:new_style}. We refer the readers to the supplemental video for more examples. We set $[E_{min}, E_{max}] = [20, 100]$ for the kinematic energy term in Equation~\ref{equ:kinematic_energy}, and $\alpha=0.12$ for the volume term in Equation~\ref{equ:volume}. In order to generate visually different styles, we deliberately loosen the spacetime bounds in Section~\ref{ssec:reward_elimination}. For example, we only bound the CoM positions, and root, ankle and neck orientations for the cartwheel.}

\subsubsection{Data-driven Styles}
\begin{figure}[t]
    \centering
    \includegraphics[width=\linewidth]{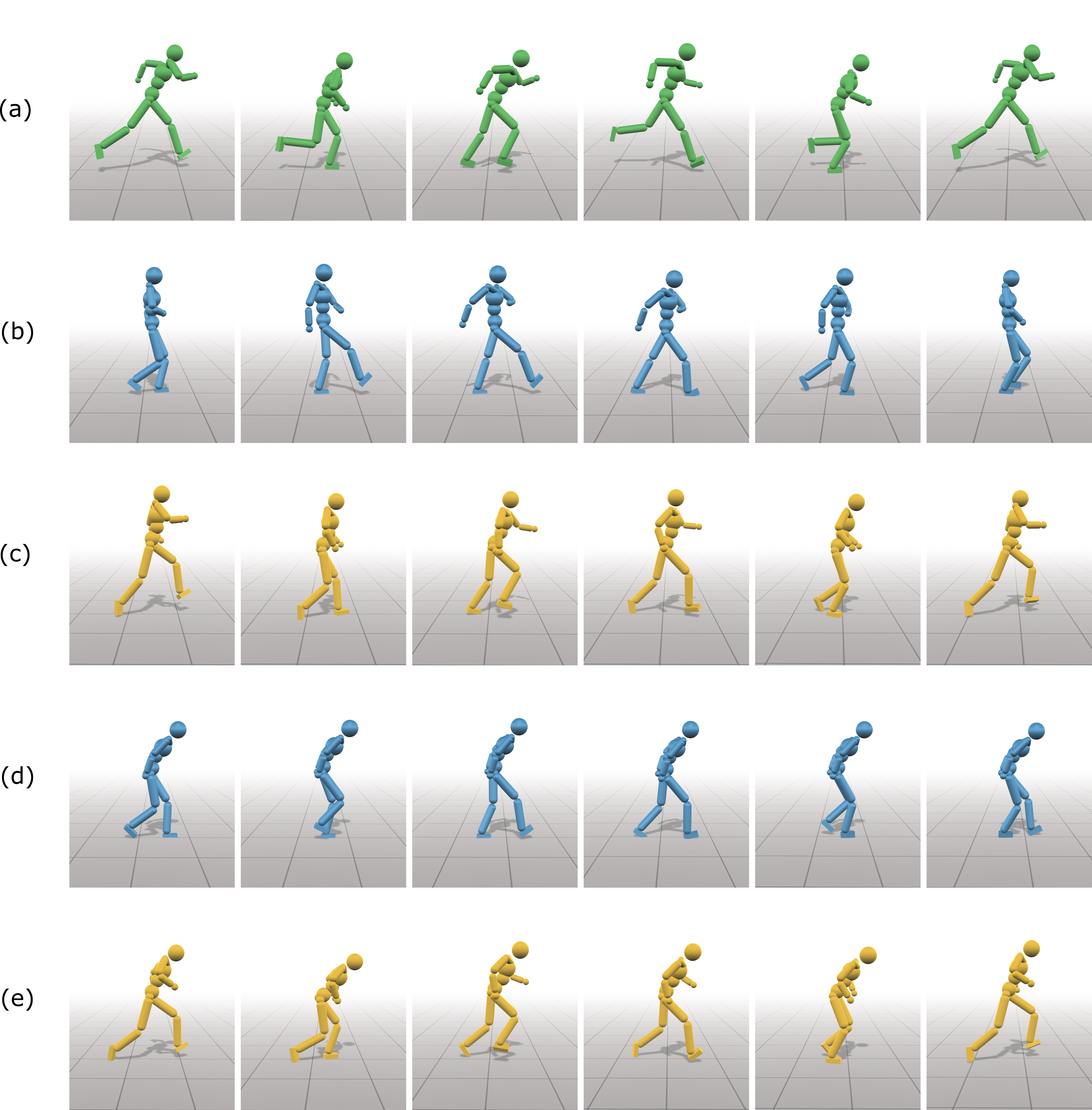}
    \caption{Style exploration with data-driven style rewards. (a) a neutral run; (b) a happy walk; (c) a happy run trained from the spacetime-bounded neutral run with the style descriptor extracted from the happy walk; (d) a bent walk; (e) a bent run trained from the spacetime-bounded neutral run with the style descriptor extracted from the bent walk.}
    \label{fig:gram_matrix}
\end{figure}

{We test style exploration using the data-driven style reward as described in Section ~\ref{sec:styleExploration} for motions selected from the CMU mocap database, as shown in Figure~\ref{fig:gram_matrix}. We directly use the autoencoder from \cite{Holden16} to encode stylistic walking motions in Figure~\ref{fig:gram_matrix}(b) and (d) to high level features for Gram matrix computation. A neutral run as shown in Figure~\ref{fig:gram_matrix}(a) is used to derive relevant spacetime bounds for DRL training. The Gram matrix for the simulated motion is computed from the current state backward in time for a fixed duration of one locomotion cycle. Then the data-driven style term can be evaluated by Equation~\ref{equ:gram_matrix} from the two Gram matrices. We again use larger spacetime bounds than those given in Section~\ref{ssec:reward_elimination} to support more aggressive style explorations for these cases.}


\begin{figure*}[t]
    \centering
    \includegraphics[width=\linewidth]{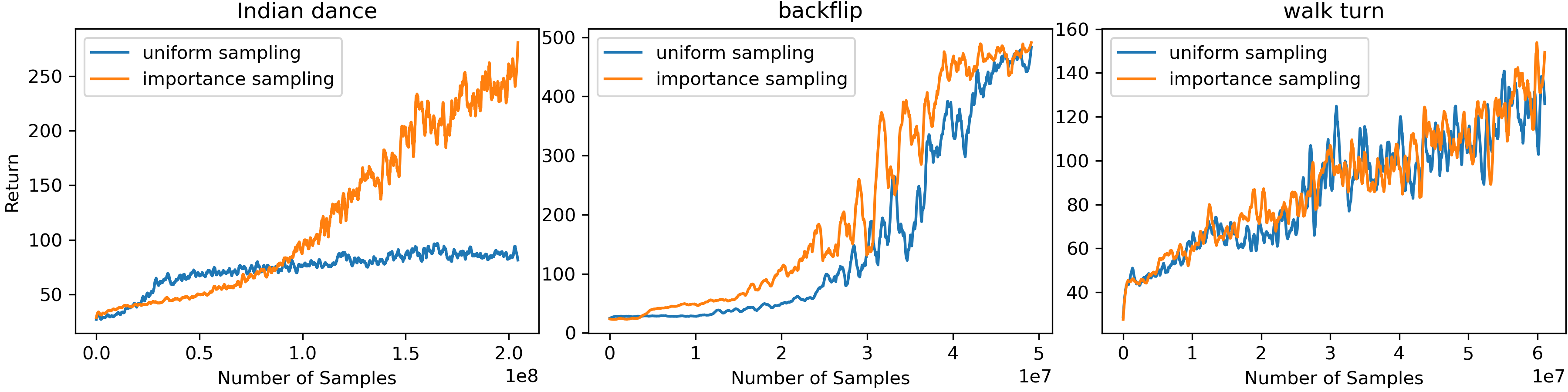}
    \caption{{Uniform vs. importance sampling for learning three skills: Indian dance, backflip and walk turn. (Left) Importance sampling is much more superior for the Indian dance, which contains a few $360^\circ$ turns. (Middle) Importance sampling is beneficial for the backflip, which contains critical points such as the takeoff. (Right) The two sampling methods do not differ much for the walk turn.}}
    \label{fig:importance_sampling}
\end{figure*}

\begin{figure}[t]
    \centering
    \includegraphics[width=\linewidth]{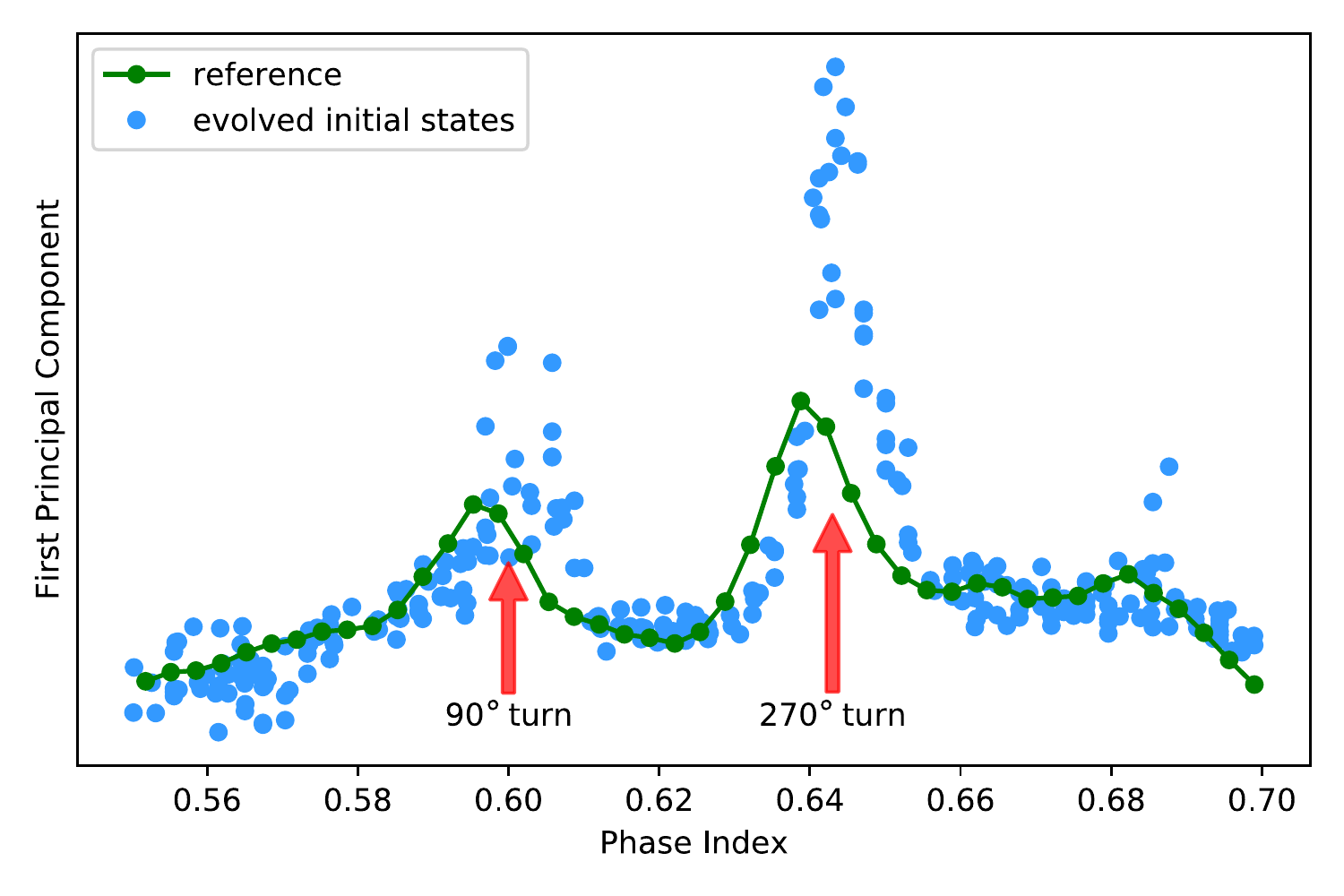}
    \caption{The first principal components of the evolved initial states (blue dots) and the reference states (green dots) for a segment of a break dance. We use the modified locally linear embedding \cite{Zhang:2007:mlle} for principal component analysis. Around sharp turns, the evolved initial states significantly deviate from the reference initial states, which enables the successful learning of this challenging skill.}
    \label{fig:evolution_states}
\end{figure}

\subsubsection{Comparison}
{We conduct comparative experiments to validate the necessity of spacetime bounds in style exploration within our DRL framework as given in Section~\ref{sec:DRLTraining}. We use a weighted average of an imitation reward term and the style reward terms as described in Section~\ref{sec:styleExploration}. That is, using a total reward defined as follows:
\begin{equation}
     r = (1 - w_s) r_i + w_s r_s,
     \label{eq:is}
\end{equation}
where $r_i$ is the imitation reward from DeepMimic~\cite{Peng:2018:DeepMimic} and $r_s$ is the style term. We test a range of $w_s$ listed in Table~\ref{tab:sensitivity}. As we can see, lower $w_s$ results in successful motor skills, but prohibits exploration of new styles. While higher $w_s$ results in either unstable or failed motor skills. {We also conduct another experiment with both the spacetime bounds and the composite imitation and style rewards in Equation~\ref{eq:is}. In such case, stylized skills can be learned for all $w_s$ without any failure. The learned skills are slightly less stylized as compared with just using the spacetime bounds and the style rewards, due to interference from the imitation term. We encourage readers to see these motions in our supplemental video.}}

\subsection{Ablation Study}
\label{ssec:ablation}
\subsubsection{Spacetime Bounds Sensitivity} 
{We analyze the sensitivity of spacetime bounds by training a series of controllers using spacetime bounds of different sizes, varying from tight to loose. For under-constrained motions with large initial feasible regions, such as walking, the learned policies change notably with respect to the size of the specified spacetime bounds. The looser the sapcetime bounds are, the more relaxed and less constrained the learned walk is. For highly constrained motions with narrow initial feasible regions, such as a cartwheel, too tight spacetime bounds result in training failures, and too loose bounds do not influence the learned skills notably. These results reveal the interactions between the spacetime bounds and the inherent feasible regions of dynamic skills. Please see the supplemental video for comparisons of the relevant animation results.} 

\subsubsection{Effect of the Feedforward Controller} 
{Integrated neural network models without separating FBC and FFC can successfully learn many motor skills \cite{Peng:2018:DeepMimic}. However, separating FFC from FBC can result in much faster learning, as proven by a few recent works \cite{Bergamin:2019:DReCon, Park:2019:LPS}. We also found that FFC helps to learn skills that contain ambiguous phase-state correspondences. For example, we demonstrate in our supplemental video that our model can successfully learn a back-bridge-with-leg-lift skill, during which the character is static for a while. In contrast, models without FFC cannot reproduce the skill at all. {The disadvantage of using FFC is that the original reference data need to be stored in a memory to compute the final policy.}}

\subsubsection{Initial States Adaptation}
{Figure~\ref{fig:importance_sampling} shows training with and without the importance sampling of reference motion as described in Section~\ref{sec:importance}. For challenging tasks such as the Indian dance, importance sampling greatly improves the learning and convergence speed. For less challenging tasks, the benefit of importance sampling will gradually diminish. Regarding the effect of initial states evolution as described in Section~\ref{sec:evolution}, we visualize the evolved initial states together with the reference states for a break dance in Figure~\ref{fig:evolution_states}. Around sharp turns of the motion, evolved initial states significantly deviate from the reference initial states, which enables the successful learning of this challenging skill. We refer readers to the supplemental video for the animation result.}

\section{Discussion}
\label{sec:discussion}


{We have presented a deep reinforcement learning framework that robustly learns motor skills via spacetime bounds. We show that our method can learn motor skills without any imitation or handcrafted rewards. Thus our method is more robust to low-quality reference motions. Moreover, spacetime bounds impose hard constraints to the training process, so the learned skills are guaranteed to be close to challenging parts of the reference skills. Furthermore, spacetime bounds can be easily combined with style exploration rewards, imitation rewards, regularization or any other mechanism such as \cite{Jiang19}, to either achieve effects such as style exploration, or to further improve synthesis quality.} 

{All the spacetime bounds that we used are derived directly from the reference motion, e.g., the target orientation plus and minus 40 degrees. The working range of spacetime bounds are usually quite large. Intuitively, when the reference quality is good, the bounds can be tighter. When the reference quality is bad, the bounds should be looser. When we just need to reproduce the reference, the bounds should be tighter. When we want to explore different styles, the bounds should be looser. Moreover, when the motion is highly constrained such as a gymnastic backflip, the size of the bounds do not affect the results that much, and therefore minimal tuning is required. For under-constrained motions such as locomotion, the size of the bounds affect the styles of the output. However, if we do not care too much about the styles, the required tuning is also minimal.} 

{We would like to note that DRL learning with imitation rewards alone already works well for reconstructing high-quality reference motions without challenging parts. However, spacetime bounds can still be used together with imitation rewards for such cases to replace ad-hoc early termination techniques such as undesired body-ground contacts. For low-quality reference motions, skills containing challenging parts, or style exploration, spacetime bounds can be used, either alone or together with imitation rewards, to lead the learning to more faithful reconstruction of the desired skills or more stylish skills. Imitation-alone methods do not work at all for style exploration. But imitation rewards can be used together with spacetime bounds and style terms, although they do interfere with style exploration to certain extent depending on the specific weighting scheme.}

{Spacetime bounds together with physical laws and system dynamics restrict and shrink the feasible region of a dynamic skill to constrain the learning in an early termination fashion. We refer interested readers to materials in mathematical physics on spacetime causal structure ~\cite{Hawking:1973:LSS,Penrose:1972:TDT,lawson:1960:relativity} to better understand the fast shrinkage of the feasible region under spacetime constraints. Currently we only use static spacetime bounds. It would be interesting to investigate how to adaptively adjust the spacetime bounds with the experiences accumulated during the learning process. Recently, powerful interactive systems that employ kinematic data-driven animation engines to train physics-based models with DRL have been developed \cite{Bergamin:2019:DReCon, Park:2019:LPS, Won20}. We also wish to integrate our stylized controllers into a more powerful system in the future, where the styles can be more explicitly activated as in~\cite{Aristidou:2017:ECU}.}

\paragraph*{Acknowledgements}
We would like to thank Xue Bin Peng and Michiel van de Panne for their suggestions on an early draft of this paper. We thank Zhiqi Yin for developing the rendering code. We also thank the anonymous reviewers for their constructive feedback. This project is partially supported by NSERC Discovery Grants Program RGPIN-06797 and RGPAS-522723. 

\bibliographystyle{eg-alpha-doi}
\bibliography{motionBound}

\newcommand{\etalchar}[1]{$^{#1}$}
\begin{thebibliography}{\uppercase{JvWdGL19}}

\bibitem[ABC96]{Amaya:1996:Emotion}
\textsc{Amaya K., Bruderlin A., Calvert T.}:
\newblock Emotion from motion.
\newblock In \emph{Graphics interface} (1996), vol.~96, pp.~222--229.

\bibitem[ABdLH13]{AlBorno13}
\textsc{Al~Borno M., de~Lasa M., Hertzmann A.}:
\newblock Trajectory optimization for full-body movements with complex
  contacts.
\newblock \emph{TVCG 19}, 8 (2013), 1405--1414.

\bibitem[ASvdP13]{Agrawal:2013:Diverse}
\textsc{Agrawal S., Shen S., van~de Panne M.}:
\newblock Diverse motion variations for physics-based character animation.
\newblock In \emph{Proceedings of the 12th ACM SIGGRAPH/Eurographics Symposium
  on Computer Animation} (2013), pp.~37--44.

\bibitem[AvdP16]{Agrawal:2016:TaskBasedLocomotion}
\textsc{Agrawal S., van~de Panne M.}:
\newblock Task-based locomotion.
\newblock \emph{ACM Transctions on Graphics 35}, 4 (2016).

\bibitem[AZS{\etalchar{*}}17]{Aristidou:2017:ECU}
\textsc{Aristidou A., Zeng Q., Stavrakis E., Yin K., Cohen-Or D., Chrysanthou
  Y., Chen B.}:
\newblock Emotion control of unstructured dance movements.
\newblock In \emph{Proceedings of the ACM SIGGRAPH/Eurographics Symposium on
  Computer Animation} (2017).

\bibitem[BCHF19]{Bergamin:2019:DReCon}
\textsc{Bergamin K., Clavet S., Holden D., Forbes J.}:
\newblock {DReCon}: data-driven responsive control of physics-based characters.
\newblock \emph{ACM Transctions on Graphics 38}, 6 (2019).

\bibitem[BH00]{Brand:2000:SM}
\textsc{Brand M., Hertzmann A.}:
\newblock Style machines.
\newblock In \emph{Proceedings of the 27th Annual Conference on Computer
  Graphics and Interactive Techniques} (2000), pp.~183--192.

\bibitem[Bul15]{Bullet}
\textsc{Bullet}:
\newblock Bullet physics library, 2015.
\newblock http://bulletphysics.org.

\bibitem[CBvdP10]{Coros10}
\textsc{Coros S., Beaudoin P., van~de Panne M.}:
\newblock Generalized biped walking control.
\newblock \emph{ACM Transctions on Graphics 29}, 4 (2010), Article 130.

\bibitem[CKJ{\etalchar{*}}11]{Coros:2011:LocomotionQuadrupeds}
\textsc{Coros S., Karpathy A., Jones B., Reveret L., van~de Panne M.}:
\newblock Locomotion skills for simulated quadrupeds.
\newblock \emph{ACM Transactions on Graphics 30}, 4 (2011).

\bibitem[Cla16]{Clavet:2016:MotionMatching}
\textsc{Clavet S.}:
\newblock Motion matching and the road to next-gen animation.
\newblock In \emph{GCD} (2016).

\bibitem[DLvdPY15]{Ding15}
\textsc{Ding K., Liu L., van~de Panne M., Yin K.}:
\newblock Learning reduced-order feedback policies for motion skills.
\newblock In \emph{Proc. ACM SIGGRAPH / Eurographics Symposium on Computer
  Animation} (2015).

\bibitem[DSAP08]{Silva:2008:ModelPredictive}
\textsc{Da~Silva M., Abe Y., Popovi{\'c} J.}:
\newblock Simulation of human motion data using short-horizon model-predictive
  control.
\newblock \emph{Computer Graphics Forum 27}, 2 (2008), 371--380.

\bibitem[Gra98]{Grassia:1998:ExpMap}
\textsc{Grassia F.~S.}:
\newblock Practical parameterization of rotations using the exponential map.
\newblock \emph{Journal of graphics tools 3}, 3 (1998), 29--48.

\bibitem[HE73]{Hawking:1973:LSS}
\textsc{Hawking S.~W., Ellis G. F.~R.}:
\newblock \emph{The large scale structure of space-time}.
\newblock Cambridge university press, 1973.

\bibitem[HKS17]{Holden:2017:PFNN}
\textsc{Holden D., Komura T., Saito J.}:
\newblock Phase-functioned neural networks for character control.
\newblock \emph{ACM Transctions on Graphics 36}, 4 (2017).

\bibitem[HPP05]{Hsu:2005:ST}
\textsc{Hsu E., Pulli K., Popovi{\'c} J.}:
\newblock Style translation for human motion.
\newblock \emph{ACM Transctions on Graphics 24}, 3 (2005), 1082--1089.

\bibitem[HRL15]{Hamalainen:2015:Online}
\textsc{H{\"a}m{\"a}l{\"a}inen P., Rajam{\"a}ki J., Liu C.~K.}:
\newblock Online control of simulated humanoids using particle belief
  propagation.
\newblock \emph{ACM Transactions on Graphics 34}, 4 (2015).

\bibitem[HSK16]{Holden16}
\textsc{Holden D., Saito J., Komura T.}:
\newblock A deep learning framework for character motion synthesis and editing.
\newblock \emph{ACM Transctions on Graphics 35}, 4 (2016), Article 138.

\bibitem[HTS{\etalchar{*}}17]{Heess:2017:Emergence}
\textsc{Heess N., TB D., Sriram S., Lemmon J., Merel J., Wayne G., Tassa Y.,
  Erez T., Wang Z., Eslami S., et~al.}:
\newblock Emergence of locomotion behaviours in rich environments.
\newblock \emph{ArXiv abs/1707.02286} (2017).

\bibitem[IAF09]{Ikemoto:2009:GME}
\textsc{Ikemoto L., Arikan O., Forsyth D.}:
\newblock Generalizing motion edits with gaussian processes.
\newblock \emph{ACM Transctions on Graphics 28}, 1 (2009).

\bibitem[JvWdGL19]{Jiang19}
\textsc{Jiang Y., van Wouwe T., de~Groote F., Liu K.~C.}:
\newblock Synthesis of biologically realistic human motion using joint torque
  actuation.
\newblock \emph{ACM Transactions on Graphics (Proc. SIGGRAPH 2019) 38}, 4
  (2019).

\bibitem[KGP02]{Kovar:2002:MotionGraphs}
\textsc{Kovar L., Gleicher M., Pighin F.}:
\newblock Motion graphs.
\newblock \emph{ACM Transctions on Graphics 21}, 3 (2002), 473--482.

\bibitem[LE60]{lawson:1960:relativity}
\textsc{Lawson R.~W., Einstein A.}:
\newblock \emph{Relativity: the special and the general theory}.
\newblock Three Rivers Press, 1960.

\bibitem[LHP{\etalchar{*}}15]{lillicrap2015continuous}
\textsc{Lillicrap T.~P., Hunt J.~J., Pritzel A., Heess N., Erez T., Tassa Y.,
  Silver D., Wierstra D.}:
\newblock Continuous control with deep reinforcement learning.
\newblock \emph{arXiv preprint arXiv:1509.02971} (2015).

\bibitem[LKL10]{LeeYS10}
\textsc{Lee Y., Kim S., Lee J.}:
\newblock Data-driven biped control.
\newblock \emph{ACM Transctions on Graphics 29}, 4 (2010), Article 129.

\bibitem[LP02]{Liu:2002:SCDC}
\textsc{Liu C.~K., Popovi\'{c} Z.}:
\newblock Synthesis of complex dynamic character motion from simple animations.
\newblock \emph{ACM Transctions on Graphics 21}, 3 (2002), 408--416.

\bibitem[LvdPY16]{Liu16}
\textsc{Liu L., van~de Panne M., Yin K.}:
\newblock Guided learning of control graphs for physics-based characters.
\newblock \emph{ACM Transctions on Graphics 35}, 3 (2016), Article 29.

\bibitem[LWB{\etalchar{*}}10]{Yongjoon:2010:MotionFields}
\textsc{Lee Y., Wampler K., Bernstein G., Popovi\'{c} J., Popovi\'{c} Z.}:
\newblock Motion fields for interactive character locomotion.
\newblock In \emph{ACM SIGGRAPH Asia 2010 Papers} (2010).

\bibitem[LWH{\etalchar{*}}12]{Levine:2012:Embedding}
\textsc{Levine S., Wang J.~M., Haraux A., Popovi\'{c} Z., Koltun V.}:
\newblock Continuous character control with low-dimensional embeddings.
\newblock \emph{ACM Transctions on Graphics 31}, 4 (2012).

\bibitem[LYG15]{Liu:2015:Samcon2}
\textsc{Liu L., Yin K., Guo B.}:
\newblock Improving sampling-based motion control.
\newblock \emph{Computer Graphics Forum 34}, 2 (2015), 415--423.

\bibitem[LYvdP{\etalchar{*}}10]{Liu:2010:Samcon}
\textsc{Liu L., Yin K., van~de Panne M., Shao T., Xu W.}:
\newblock Sampling-based contact-rich motion control.
\newblock \emph{ACM Transctions on Graphics 29}, 4 (2010).

\bibitem[LYvdPG12]{liu2012terrain}
\textsc{Liu L., Yin K., van~de Panne M., Guo B.}:
\newblock Terrain runner: control, parameterization, composition, and planning
  for highly dynamic motions.
\newblock \emph{ACM Transctions on Graphics 31}, 6 (2012), 154.

\bibitem[MLC10]{Min:2010:SEP}
\textsc{Min J., Liu H., Chai J.}:
\newblock Synthesis and editing of personalized stylistic human motion.
\newblock In \emph{Proceedings of the 2010 ACM SIGGRAPH symposium on
  Interactive 3D Graphics and Games} (2010), pp.~39--46.

\bibitem[MYGY19]{Ma:2019:TRD}
\textsc{Ma L.-K., Yang Z., Guo B., Yin K.}:
\newblock Towards robust direction invariance in character animation.
\newblock \emph{Computer Graphics Forum 38}, 7 (2019).

\bibitem[PALvdP18]{Peng:2018:DeepMimic}
\textsc{Peng X.~B., Abbeel P., Levine S., van~de Panne M.}:
\newblock {DeepMimic}: Example-guided deep reinforcement learning of
  physics-based character skills.
\newblock \emph{ACM Transctions on Graphics 37}, 4 (2018).

\bibitem[PBYvdP17]{2017-TOG-deepLoco}
\textsc{Peng X.~B., Berseth G., Yin K., van~de Panne M.}:
\newblock {DeepLoco}: Dynamic locomotion skills using hierarchical deep
  reinforcement learning.
\newblock \emph{ACM Transactions on Graphics (Proc. SIGGRAPH 2017) 36}, 4
  (2017).

\bibitem[Pen87]{Penrose:1972:TDT}
\textsc{Penrose R.}:
\newblock \emph{Techniques of differential topology in relativity}.
\newblock Society for Industrial and Applied Mathematics, 1987.

\bibitem[PKM{\etalchar{*}}18]{Peng:2018:SFV}
\textsc{Peng X.~B., Kanazawa A., Malik J., Abbeel P., Levine S.}:
\newblock {SFV}: Reinforcement learning of physical skills from videos.
\newblock \emph{ACM Transctions on Graphics 37}, 6 (2018).

\bibitem[PRL{\etalchar{*}}19]{Park:2019:LPS}
\textsc{Park S., Ryu H., Lee S., Lee S., Lee J.}:
\newblock Learning predict-and-simulate policies from unorganized human motion
  data.
\newblock \emph{ACM Transctions on Graphics 38}, 6 (2019).

\bibitem[PyT18]{pytorch}
\textsc{PyTorch}:
\newblock Pytorch, 2018.
\newblock https://pytorch.org/.

\bibitem[SB18]{Sutton:2018:reinforcement}
\textsc{Sutton R.~S., Barto A.~G.}:
\newblock \emph{Reinforcement learning: An introduction}.
\newblock MIT press, 2018.

\bibitem[SCF06]{Shapiro:2006:SC}
\textsc{Shapiro A., Cao Y., Faloutsos P.}:
\newblock Style components.
\newblock In \emph{Proceedings of Graphics Interface 2006} (2006), pp.~33--39.

\bibitem[SH07]{Safonova:2007:OptimalSearch}
\textsc{Safonova A., Hodgins J.~K.}:
\newblock Construction and optimal search of interpolated motion graphs.
\newblock \emph{ACM Transctions on Graphics 26}, 3 (2007), 106--es.

\bibitem[SHP04]{Safonova:2004:OptPCA}
\textsc{Safonova A., Hodgins J.~K., Pollard N.~S.}:
\newblock Synthesizing physically realistic human motion in low-dimensional,
  behavior-specific spaces.
\newblock \emph{ACM Transctions on Graphics 23}, 3 (2004), 514--521.

\bibitem[SML{\etalchar{*}}15]{Schulman:2015:GAE}
\textsc{Schulman J., Moritz P., Levine S., Jordan M., Abbeel P.}:
\newblock High-dimensional continuous control using generalized advantage
  estimation.
\newblock \emph{CoRR abs/1506.02438} (2015).

\bibitem[SWD{\etalchar{*}}17]{Schulman:2017:PPO}
\textsc{Schulman J., Wolski F., Dhariwal P., Radford A., Klimov O.}:
\newblock Proximal policy optimization algorithms.
\newblock \emph{CoRR abs/1707.06347} (2017).

\bibitem[TLT11]{Tan:2011:SPD}
\textsc{Tan J., Liu K., Turk G.}:
\newblock Stable proportional-derivative controllers.
\newblock \emph{IEEE Computer Graphics and Applications 31}, 4 (2011), 34--44.

\bibitem[WFH07]{Wang:2007:MGP}
\textsc{Wang J.~M., Fleet D.~J., Hertzmann A.}:
\newblock Multifactor gaussian process models for style-content separation.
\newblock In \emph{Proceedings of the 24th International Conference on Machine
  Learning} (2007), pp.~975--982.

\bibitem[WGH20]{Won20}
\textsc{Won J., Gopinath D., Hodgins J.~K.}:
\newblock A scalable approach to control diverse behaviors for physically
  simulated characters.
\newblock \emph{ACM Transctions on Graphics 39}, 4 (2020).

\bibitem[WL19]{won2019learning}
\textsc{Won J., Lee J.}:
\newblock Learning body shape variation in physics-based characters.
\newblock \emph{ACM Transactions on Graphics (TOG) 38}, 6 (2019), Article 207.

\bibitem[XLKvdP20]{xie2020allsteps}
\textsc{Xie Z., Ling H.~Y., Kim N.~H., van~de Panne M.}:
\newblock Allsteps: Curriculum-driven learning of stepping stone skills.
\newblock In \emph{Proceedings of the ACM SIGGRAPH/Eurographics Symposium on
  Computer Animation} (2020).

\bibitem[XWCH15]{Xia:2015:RST}
\textsc{Xia S., Wang C., Chai J., Hodgins J.}:
\newblock Realtime style transfer for unlabeled heterogeneous human motion.
\newblock \emph{ACM Transctions on Graphics 34}, 4 (2015).

\bibitem[YCP03]{Yin03}
\textsc{Yin K., Cline M.~B., Pai D.~K.}:
\newblock Motion perturbation based on simple neuromotor control models.
\newblock In \emph{Proceedings of Pacific Graphics} (2003).

\bibitem[YL10]{Ye10}
\textsc{Ye Y., Liu C.~K.}:
\newblock Optimal feedback control for character animation using an abstract
  model.
\newblock \emph{ACM Transctions on Graphics 29}, 4 (2010), Article 74.

\bibitem[YLvdP07]{Yin07}
\textsc{Yin K., Loken K., van~de Panne M.}:
\newblock {SIMBICON}: Simple biped locomotion control.
\newblock \emph{ACM Transctions on Graphics 26}, 3 (2007), Article 105.

\bibitem[YM16]{Yumer:2016:spectral}
\textsc{Yumer M.~E., Mitra N.~J.}:
\newblock Spectral style transfer for human motion between independent actions.
\newblock \emph{ACM Transctions on Graphics 35}, 4 (2016).

\bibitem[YTL18]{Yu:2018:LSLL}
\textsc{Yu W., Turk G., Liu C.~K.}:
\newblock Learning symmetric and low-energy locomotion.
\newblock \emph{ACM Transactions on Graphics 37}, 4 (2018).

\bibitem[ZSKS18]{Zhang:2018:MANN}
\textsc{Zhang H., Starke S., Komura T., Saito J.}:
\newblock Mode-adaptive neural networks for quadruped motion control.
\newblock \emph{ACM Transctions on Graphics 37}, 4 (2018).

\bibitem[ZW07]{Zhang:2007:mlle}
\textsc{Zhang Z., Wang J.}:
\newblock {MLLE}: Modified locally linear embedding using multiple weights.
\newblock In \emph{Advances in neural information processing systems} (2007),
  pp.~1593--1600.

\end{thebibliography}

\end{document}